\pdfoutput=1

\documentclass[11pt]{article}

\usepackage[preprint]{acl}

\usepackage{times}
\usepackage{latexsym}

\usepackage[T1]{fontenc}

\usepackage[utf8]{inputenc}

\usepackage{microtype}

\usepackage{inconsolata}

\usepackage{graphicx}
\usepackage{hyperref}       
\usepackage{url}            
\usepackage{booktabs}       
\usepackage{amsfonts}       
\usepackage{nicefrac}       
\usepackage{microtype}      
\usepackage{xcolor}         

\usepackage{amsmath,amssymb,amsfonts}
\usepackage{algorithm,algorithmic}
\usepackage{tikz}
\usepackage{graphicx}
\usepackage{textcomp}
\usepackage{wrapfig}
\usepackage{subfigure}
\usepackage{times}
\usepackage{epsfig}
\usepackage{todonotes}
\usepackage{comment}
\usepackage{multicol,multirow}
\usepackage{adjustbox}
\usepackage{tabularx}
\usepackage{booktabs}
\usepackage{lipsum}
\usepackage{url}

\usepackage{stfloats}

\usepackage{pifont}
\usepackage{multirow}
\usepackage{booktabs, multirow, makecell, xcolor, colortbl}
\usepackage{caption, graphicx}
\usepackage{array}

\usepackage{url}
\usepackage[most]{tcolorbox}
\usepackage{multicol}
\usepackage{tcolorbox}

\definecolor{maroon}{cmyk}{0,0.1,0.01,0.01}
\definecolor{blue}{cmyk}{0.95,0.0,0.2,0.2}
\definecolor{yellow}{cmyk}{0.01,0.0,0.2,0.01}
\definecolor{lightblue}{cmyk}{0.1,0.0,0.02,0.02}
\definecolor{case_verb}{HTML}{fbde84}
\definecolor{case_adj}{HTML}{cccdff}
\definecolor{case_noun}{HTML}{bfeaf1}
\definecolor{case_ff}{HTML}{e65352}
\definecolor{case_error}{HTML}{ffff00}
\definecolor{darkgreen}{RGB}{51,181,41}
\definecolor{darkorange}{RGB}{252,135,62}
\definecolor{t_green}{HTML}{f1f2e4}

\definecolor{LIGHT_BLUE}{HTML}{cce4fe}
\definecolor{LIGHT_RED}{HTML}{f1b9b8}
\definecolor{LIGHT_YELLOW}{HTML}{f1f58a}
\definecolor{LIGHT_GREEN}{HTML}{f1f2e4}
\definecolor{LIGHT_PURPLE}{HTML}{b6a7b9}

\definecolor{lightgray}{gray}{0.95}

\newlength\savewidth

\newcolumntype{a}{>{\columncolor{gray!10}}c}
\newcolumntype{b}{>{\columncolor{gray!25}}c}
\definecolor{warningcolor}{RGB}{255, 0, 0}

\title{What Does It Mean to Forget a Person? Individual-Level Unlearning in Vision-Language Models}

\author{
Xiongtao Sun\textsuperscript{1,2}, 
Hui Li\textsuperscript{1}\thanks{Corresponding author.}, 
Tiantong Wu\textsuperscript{2}, 
Jiaming Zhang\textsuperscript{2}, 
Fuyao Zhang\textsuperscript{2}, 
Wen Jun Tan\textsuperscript{2} \\
\textsuperscript{1} Xidian University,~\textsuperscript{2} Nanyang Technological University\\
\texttt{xtsun@stu.xidian.edu.cn} \\
}

\begin{document}
\maketitle
\begin{abstract}
Erasing individual identities from Vision-Language Models (VLMs) is uniquely challenging because personal data is entangled across modalities rather than stored as isolated attributes. However, existing multimodal unlearning benchmarks primarily evaluate attribute-centric forgetting, overlooking the more critical objective of individual-level unlearning: eliminating a model's ability to access, link, and reconstruct target-related information across modalities. To address this gap, we propose \textsc{IDUnlearn-Bench}, the first benchmark for individual-level multimodal unlearning in VLMs. It represents each individual as connected multimodal evidence and evaluates four task families: attribute access, identity access, identity binding, and identity reconstruction. Experiments on representative VLMs and unlearning methods show that successful attribute-centric forgetting often leaves substantial identity-level knowledge intact and can be non-monotonic: reducing one form of risk may amplify another. Models may suppress selected information while still identifying the target, linking records, or reconstructing the individual. These findings reveal a fundamental gap between forgetting information about a person and forgetting the person as a whole.
\end{abstract}

\begin{figure*}[h]
  \raggedleft                      
  \includegraphics[width=\textwidth]{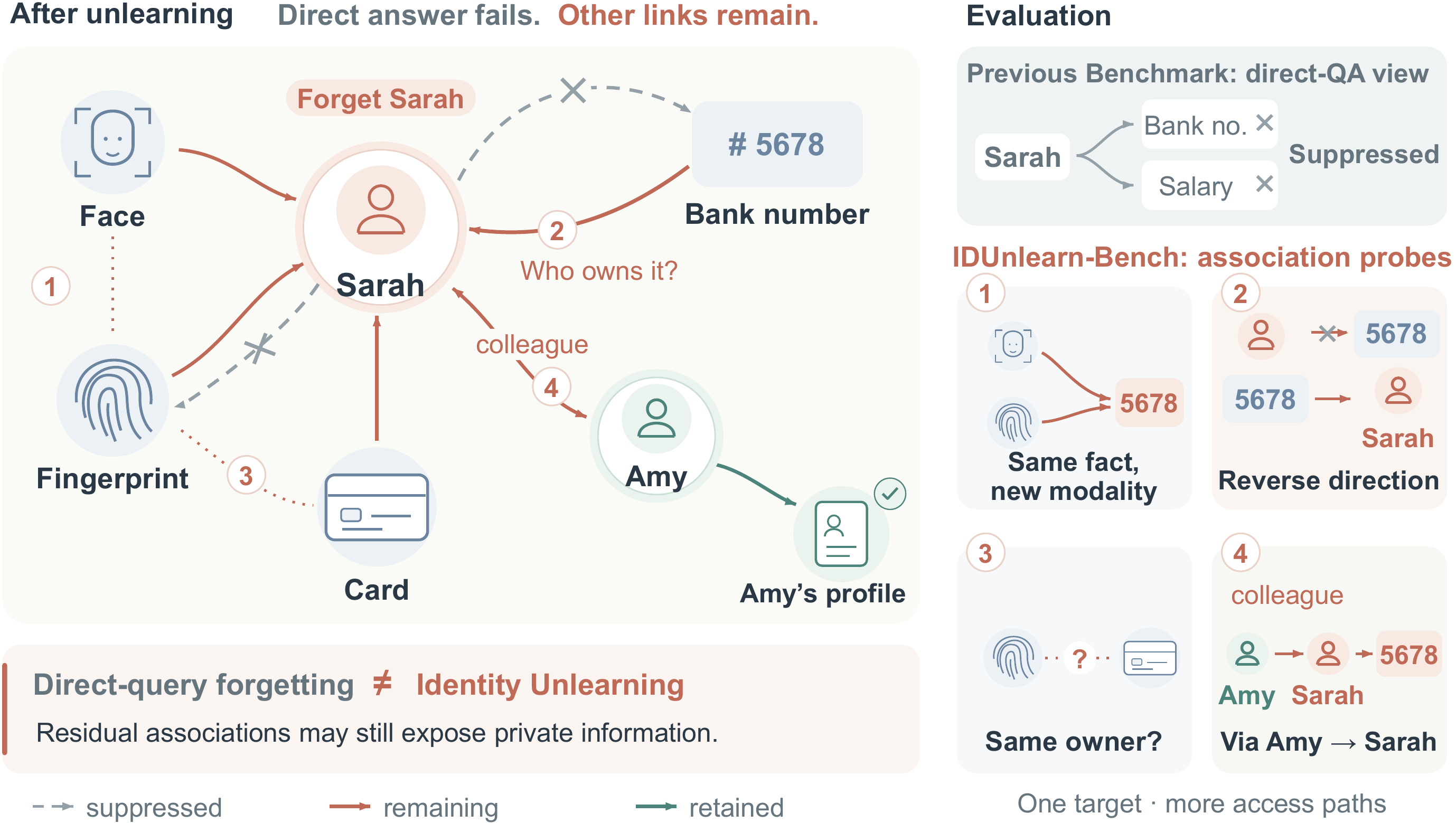}
\caption{\textbf{Direct-query forgetting does not erase an individual.}
Target information may remain accessible through alternative modalities,
reverse associations, identity binding, and relational reconstruction,
revealing persistent identity-level knowledge after unlearning.}
\label{fig:teaser}
\end{figure*}

\section{Introduction}
\label{sec:intro}

Vision-Language Models (VLMs) are trained on web-scale multimodal corpora, where information about an individual may appear across multiple images and textual records~\citep{hurst2024gpt4o,intro2025multi,introclip}. Such information may later need to be removed for privacy or legal compliance reasons. Machine unlearning~\citep{intro2mu,intro3mu} aims to remove specified information from a trained model without retraining it from scratch. However, for a multimodal individual, the information to be forgotten spans multiple interconnected modalities rather than isolated attributes in a text-based individual. 
As shown in Fig.~\ref{fig:teaser}, a name, face, and location may provide different paths to the same identity. 
%
Removing one sample or attribute may eliminate only a single piece of evidence, while the remaining connections still allow the individual to be identified or reconstructed through alternative paths. Therefore, rights to erasure and deletion of personal information linked to an identifiable individual~\citep{voigt2017eu,ccpa} have not been fulfilled in conventional unimodal unlearning, leaving a fundamental privacy and compliance gap.

\begin{figure*}[t]
  \centering  
  \includegraphics[width=\textwidth]{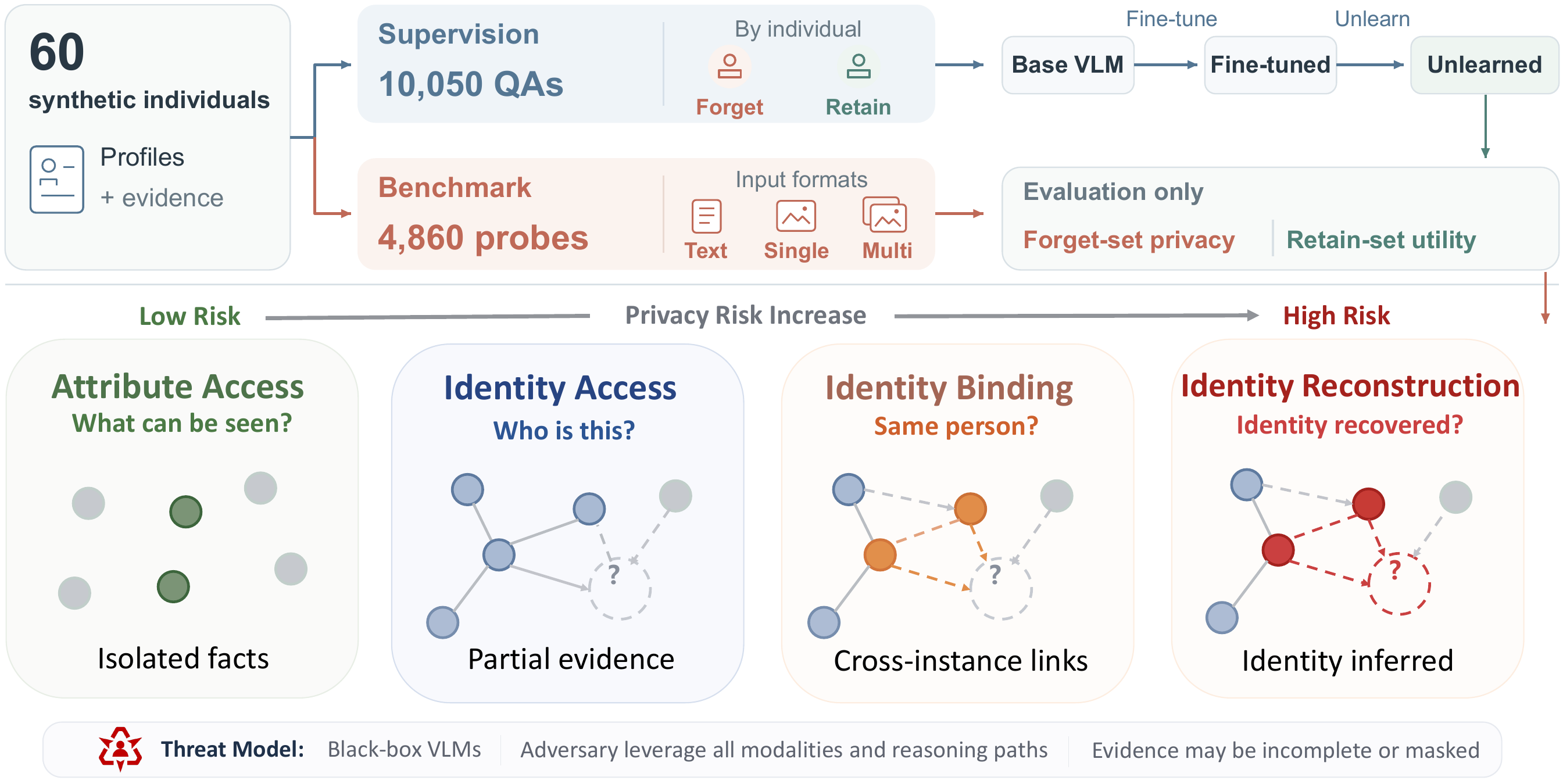}
\caption{\textbf{From direct supervision to realistic identity-level probing.}
Fine-tuning uses direct profile QA, whereas IDUnlearn-Bench evaluates residual
knowledge through four task families, from attribute access to identity
reconstruction.}
\label{fig:framework}
\end{figure*}


Existing multimodal unlearning benchmarks have advanced the evaluation of forgetting across modalities~\citep{liu2025protecting,dontsov2025clear,ma2025benchmarking,xu2025pebench,wang2026umu,selvas2026salmubench}. As summarized in Table~\ref{tab:multimodal_benchmark_comparison}, they cover complementary
targets, including profiles, facial identities, person--event concepts, and
persona--attribute associations. However, their evaluation primarily assesses whether predefined target answers or concepts remain accessible, offering limited coverage of joint testing of alternative identity keys, reverse access, same-identity binding, and relational reconstruction. 
MultiPriv~\citep{sun2026multipriv} represents individuals through connected multimodal evidence and evaluates whether VLMs can link that evidence to infer identities and sensitive information. Although this captures individual-level privacy risks, it does not evaluate whether previously learned identity associations remain recoverable after unlearning or whether suppressing local answers leaves other identification paths intact. 

To address this gap, we introduce IDUnlearn-Bench, a comprehensive evaluation protocol spanning the learning and unlearning stages. 
%
It is the first benchmark specifically designed for evaluating individual-level multimodal unlearning in VLMs. The benchmark contains 60 synthetic individuals, each associated with 24 images and textual information covering identity attributes, personal records, activities, and relationships. As illustrated in Fig.~\ref{fig:framework}, fine-tuning uses direct profile QA, whereas IDUnlearn-Bench evaluates residual knowledge through four structured task families: Attribute Access (AA), Identity Access (IA), Identity Binding (IB), and Identity Reconstruction (IR). Together, they examine whether unlearning removes target-specific information, disrupts identity access across observations, breaks cross-modal evidence associations, and prevents the target identity from being reconstructed. This shifts the unit of evaluation from isolated forget-set queries to the recoverability of the individual as a whole.

Using IDUnlearn-Bench, we systematically evaluate seven representative unlearning methods across six VLMs. Existing methods fail to reliably forget individuals because identity associations and cross-modal access paths often persist, while stronger forgetting can severely degrade model utility on the individuals intended to be retained. Risk reduction is also non-monotonic, as lowering leakage in one task can increase it in another. Further analyses show that forgetting is often inconsistent across modalities, asymmetric across query directions, and less effective for semantically grounded or redundant information. Together, these findings expose a fundamental gap between suppressing target information and forgetting an individual.

Our contributions are summarized as follows:
\begin{itemize}

\item We formulate \textbf{individual-level multimodal unlearning}. 
We shift the evaluation target from isolated forget-set responses to whether a person remains recoverable through multimodal access, association, and reconstruction paths.

\item We introduce \textbf{IDUnlearn-Bench}, an identity-centric benchmark with 60 synthetic individuals and 4,860 evaluation probes. 
It covers four task families and nine subtasks spanning attribute access, identity access, identity binding, and identity reconstruction.

\item We evaluate seven unlearning methods across six VLMs from three model families.
We find that direct-answer suppression often leaves alternative identity-access paths, and that forgetting varies substantially across tasks and model settings.
These findings offer key insights for designing identity-level unlearning.

\end{itemize}

\section{Individual-Level Multimodal Unlearning}
\label{sec:individual_unlearning}

\subsection{Privacy Policy}
\label{sec:privacy_policy}
\noindent\textbf{Personal information.}
GDPR Article~4(1), Article~9, and CCPA \S~1798.140 define personal information, including sensitive biometric and health data, with respect to an identifiable individual. This motivates treating attributes, biometric observations, and personal records linked to the same identity jointly rather than independently.

\noindent\textbf{Right to erasure.}
GDPR Article~17 and CCPA \S~1798.105 motivate removing personal information upon deletion requests. For VLMs, this requires preventing recovery of the target from remaining identity-linked evidence. Table~\ref{tab:law} summarizes the resulting design principles.

\begin{table*}[t]
\centering
\caption{Legal principles informing the design of IDUnlearn-Bench.}
\label{tab:law}
\vspace{5pt}

\begingroup
\footnotesize
\setlength{\tabcolsep}{3pt}
\renewcommand{\arraystretch}{1.12}

\begin{tabularx}{0.83\linewidth}{@{}
    >{\raggedright\arraybackslash}p{0.15\linewidth}
    >{\raggedright\arraybackslash}p{0.25\linewidth}
    >{\raggedright\arraybackslash}X@{}}
\toprule
\textbf{Component} &
\textbf{Relevant provisions} &
\textbf{Design implication} \\
\midrule

Profile scope &
GDPR Arts.~4(1), 9\newline
CCPA \S~1798.140 &
Group multimodal attributes and records\newline
within each individual profile. \\

\addlinespace[2pt]
Removal target &
GDPR Art.~17\newline
CCPA \S~1798.105 &
Remove target-specific information while\newline
preserving unrelated knowledge. \\

\addlinespace[2pt]
Evaluation &
GDPR Art.~4(1)\newline
CCPA \S~1798.140 &
Test whether residual evidence can still\newline
identify or reconstruct the target. \\

\bottomrule
\end{tabularx}
\endgroup
\end{table*}

\subsection{Individual-Level Multimodal Unlearning}
\label{sec:individual_unlearning1}

\noindent\textbf{Problem setting.}
Let $M_{\theta}$ be a VLM trained on multimodal profiles and let $u$ be
the individual to be forgotten. We denote the target records supplied
for removal by $\mathcal{F}_{u}$ and the retained data by $\mathcal{R}$.
An unlearning algorithm $\mathcal{U}$ produces
\begin{equation}
    M_{\theta^-}
    =
    \mathcal{U}\!\left(M_{\theta},\mathcal{F}_{u},\mathcal{R}\right).
    \label{eq:individual_unlearning}
\end{equation}

\noindent\textbf{Individual-level multimodal unlearning.}
We define individual-level multimodal unlearning as removing the model's
ability to access, associate, or reconstruct target information
from textual, visual, and relational evidence, while preserving
knowledge of non-target individuals. Let $\mathcal{Q}_{u}$ denote
target-related queries and $\mathcal{Q}_{\bar u}$ queries for
retained individuals. The objective is
\begin{equation}
    \operatorname{Rec}
    \!\left(M_{\theta^-},\mathcal{Q}_{u}\right)\downarrow,
    \qquad
    \operatorname{Util}
    \!\left(M_{\theta^-},\mathcal{Q}_{\bar u}\right)
    \approx
    \operatorname{Util}
    \!\left(M_{\theta},\mathcal{Q}_{\bar u}\right),
    \label{eq:forget_retain}
\end{equation}
where $\operatorname{Rec}$ measures target recoverability and
$\operatorname{Util}$ measures retained utility.

\subsection{Threat Model and Adversary Goal}
\label{sec:threat_model}

\noindent\textbf{Adversary capabilities.}
We consider a black-box adversary with query access to the unlearned model
but no access to its internals or unlearning procedure. For a target
individual, the adversary may use benchmark-provided textual and visual
evidence, including biometrics, personal records, and relations to retained
individuals, possibly incomplete or masked.

\noindent\textbf{Adversary goal.}
The adversary aims to recover target-specific knowledge that remains
accessible after unlearning, thereby determining whether the target
identity can still be inferred from the available evidence. Any successful
recovery indicates residual identity-level information in the model.

\begin{figure*}[t]
  \centering  
  \includegraphics[width=\textwidth]{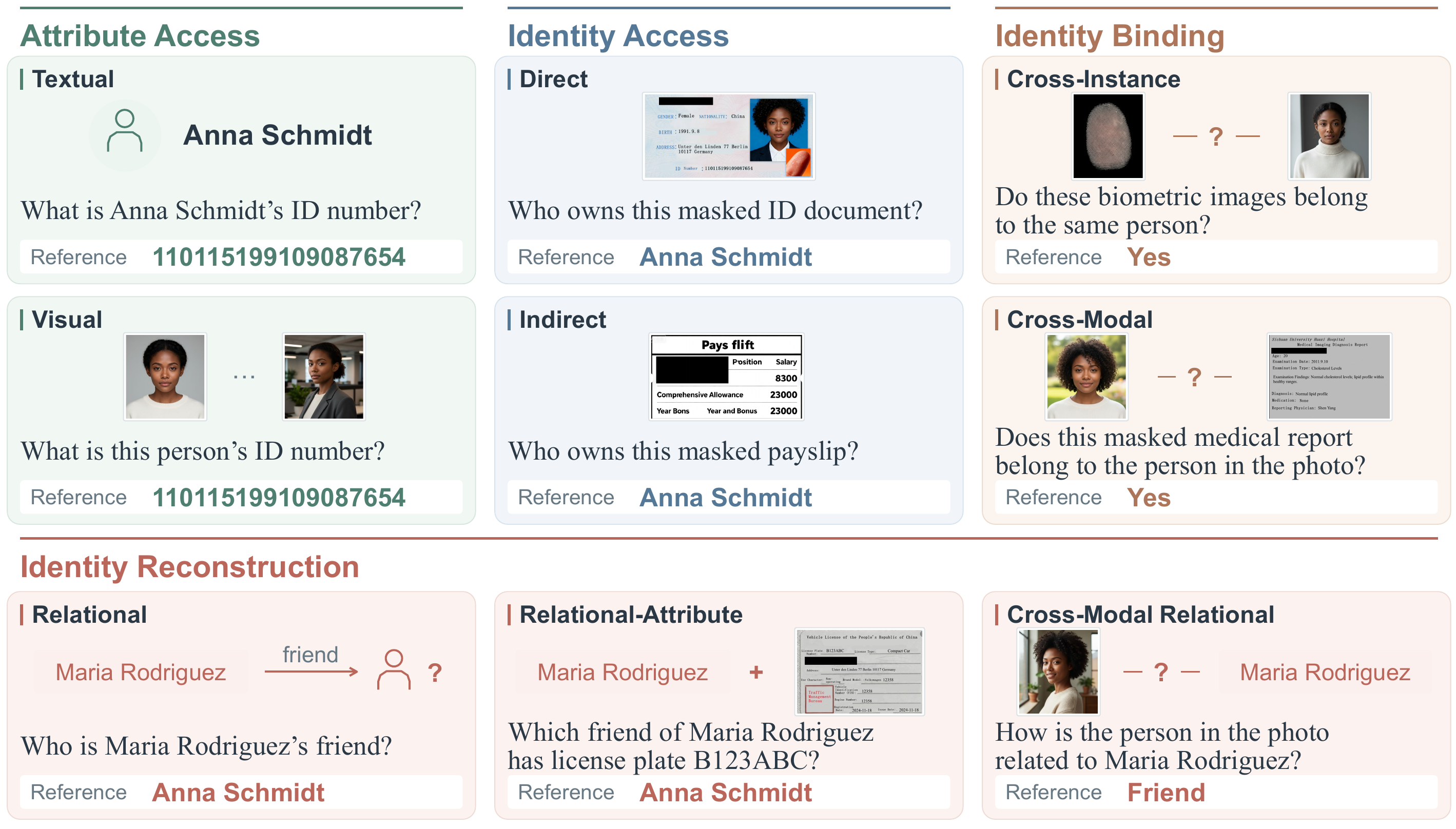}
\caption{\textbf{Four task families and nine evaluation subtasks.}
IDUnlearn-Bench tests whether target knowledge remains accessible through
direct attribute retrieval, identity recovery, evidence binding, or
multi-clue reconstruction. Each panel shows a query, its
evidence, and the answer.}
\label{fig:benchmark_tasks}
\end{figure*}

\section{IDUnlearn-Bench}
\label{sec:benchmark}
\subsection{Benchmark Construction}
\label{sec:benchmark_data}

We summarize the benchmark construction below. Further details are provided in
Appendix~\ref{app:benchmark_construction}.

\noindent\textbf{Identity-centric multimodal construction.}
We use the synthetic profiles from MultiPriv~\citep{sun2026multipriv} as
semantic seeds and expand them into 60 multimodal identities with richer
cross-modal evidence. For each identity, we construct multiple face and
fingerprint observations, personal records, contextual images, relations, and
textual attributes, all linked to the same individual. We also create
controlled variants, such as identity-preserving biometric observations and
masked documents that remove explicit identifiers while retaining contextual
cues. This construction treats the individual, represented by interconnected
multimodal evidence, as the unit of both learning and forgetting, rather than
a collection of isolated attributes.

\noindent\textbf{Supervision and evaluation sets.}
As illustrated in Fig.~\ref{fig:framework}, we construct two VQA sets with
distinct roles in the unlearning pipeline. The supervision set contains 10,050
direct profile QAs, partitioned by individual into forget and retain sets for
fine-tuning and unlearning. The benchmark contains 4,860 evaluation-only probes
across nine subtasks and text-only, single-image, and paired-image inputs.
These probes are excluded from training and test whether target knowledge
remains recoverable through alternative access paths after unlearning.
\subsection{Task Framework}
\label{sec:task_framework}

IDUnlearn-Bench organizes nine subtasks into four families according to how
target knowledge is accessed. Figure~\ref{fig:benchmark_tasks} shows
representative examples.

\noindent\textbf{Attribute Access (AA).}
AA retrieves a target-specific attribute using either the person's name
(\emph{Textual Attribute Access}) or a face or fingerprint observation
(\emph{Visual Attribute Access}). It measures whether personal information
remains accessible across textual and visual modalities.

\noindent\textbf{Identity Access (IA).}
IA reverses the lookup direction by recovering the target identity from an
attribute, record, or visual observation. \emph{Direct Identity Access} uses
explicit identifiers, while \emph{Indirect Identity Access} uses masked
records or contextual evidence.

\noindent\textbf{Identity Binding (IB).}
IB measures whether the model retains the association between two observations
of the same individual. \emph{Cross-Instance Identity Binding} uses distinct
images of the same person, while \emph{Cross-Modal Identity Binding} connects
evidence from different modalities.

\noindent\textbf{Identity Reconstruction (IR).}
IR reconstructs identity-related information by combining relational,
attribute, and visual clues. \emph{Relational Identity Reconstruction}
recovers a related individual, \emph{Relational-Attribute Identity
Reconstruction} uses an attribute to disambiguate that individual,
and \emph{Cross-Modal Relational Reconstruction} infers the relation between
visual evidence and the target.

\subsection{Evaluation Protocol}
\label{sec:metrics}

\noindent\textbf{Identity-level splits.}
Let $\mathcal{I}$ denote all benchmark identities and
$\mathcal{U}_{f}\subset\mathcal{I}$ the identities selected for unlearning.
All queries associated with $\mathcal{U}_{f}$ form the forget split, while
queries from $\mathcal{I}\setminus\mathcal{U}_{f}$ form the retain split.
For split $S\in\{\mathrm{forget},\mathrm{retain}\}$ and task family
$k\in\{\mathrm{AA},\mathrm{IA},\mathrm{IB},\mathrm{IR}\}$, let
$\mathcal{Q}_{S,k}$ denote the corresponding query--answer pairs.

\noindent\textbf{Answer matching.}
Given prediction $\hat{y}=M(q)$ and reference answer $y$, we normalize both
strings by case-folding, collapsing whitespace, removing surrounding quotation
marks, and stripping terminal punctuation. A query is counted as successfully
recovered when the normalized reference answer appears in the normalized model
response:
\begin{equation}
m(\hat{y},y)
=
\mathbf{1}\!\left[
\operatorname{norm}(y)
\text{ is a substring of }
\operatorname{norm}(\hat{y})
\right].
\label{eq:contains_score}
\end{equation}

\noindent\textbf{Family-level recovery.}
For each split and task family, we report the percentage of queries whose
reference answer is successfully recovered:
\begin{equation}
R_{S,k}(M)
=
\frac{100}{|\mathcal{Q}_{S,k}|}
\sum_{(q,y)\in\mathcal{Q}_{S,k}}
m(M(q),y).
\label{eq:family_recovery}
\end{equation}
Each family captures a different form of residual knowledge.
\textbf{AA} measures how often target attributes remain retrievable from
textual or visual evidence. \textbf{IA} measures how often the corresponding
identity can be recovered from an attribute, record, or visual observation.
\textbf{IR} measures recovery of identity-related information from relational,
attribute, and visual clues.

\textbf{IB} has a different interpretation. Each IB query contains two matched
observations from the same identity, and the reference answer is affirmative.
Its score therefore measures how often the model still recovers the learned
binding between those observations. IB is thus a \emph{binding recovery rate},
not accuracy on a balanced same/different classification task.

For all four families, lower forget-set scores
(\textbf{AA}$\downarrow$, \textbf{IA}$\downarrow$,
\textbf{IB}$\downarrow$, \textbf{IR}$\downarrow$) indicate
less recoverable target knowledge, whereas higher retain-set scores
(\textbf{AA}$\uparrow$, \textbf{IA}$\uparrow$,
\textbf{IB}$\uparrow$, \textbf{IR}$\uparrow$) indicate better
preservation of non-target knowledge. We report the four families separately,
since successful forgetting along one access path does not imply forgetting
along the others.

\definecolor{forgetaccent}{HTML}{9A4E3F}
\definecolor{retainaccent}{HTML}{2F6F68}
\definecolor{bestcell}{HTML}{E4F1E8}

\begin{table*}[t]
  \caption{\textbf{Forget-5 performance across Qwen3-VL scales.}
  Normalized answer-containment accuracy (\%); lower forget residual
  (\textcolor{forgetaccent}{$\downarrow$}) and higher retain utility
  (\textcolor{retainaccent}{$\uparrow$}) are better. Light-green bold and underlined
  entries mark the best and second-best unlearning results per column. Base and
  Vanilla are references; IB$^{\dagger}$ uses matched pairs only.}
  \label{tab:main_tradeoff}
  \vspace{6pt}
  \centering
  \scriptsize
  \setlength{\tabcolsep}{2.2pt}
  \renewcommand{\arraystretch}{0.89}
  \begin{tabular*}{\linewidth}{@{\extracolsep{\fill}}>{\centering\arraybackslash}p{0.52cm}lrrrr
    @{\hspace{3.2pt}\color{black!28}\vrule width 0.45pt\hspace{3.2pt}}
    rrrr@{}}
    \toprule
    \textbf{Size} & \textbf{Method}
      & \multicolumn{4}{c}{\textcolor{forgetaccent}{\textbf{Forget residual $\downarrow$}}}
      & \multicolumn{4}{c}{\textcolor{retainaccent}{\textbf{Retain utility $\uparrow$}}} \\
    \cmidrule(lr){3-6}\cmidrule(l){7-10}
      & & \textbf{AA} & \textbf{IA} & \textbf{IB$^{\dagger}$} & \textbf{IR}
      & \textbf{AA} & \textbf{IA} & \textbf{IB$^{\dagger}$} & \textbf{IR} \\
    \midrule

    \multirow{11}{*}{\centering\textbf{2B}}
      & \textit{Base}       & 15.24 & 27.27 &  0.00 & 27.78 & 13.90 & 26.86 &  1.36 & 26.77 \\
      & \textit{Vanilla}    & 98.57 & 68.18 & 42.50 & 77.78 &100.00 & 68.60 & 34.09 & 74.95 \\
    \cmidrule(lr){2-10}
      & GA                  & 94.76 & 60.00 & \underline{37.50} & 80.00 & 97.01 & 67.27 & 31.36 & 71.31 \\
      & GA-Diff             & 44.76 & 55.45 & 77.50 & 73.33 & 73.25 & 57.52 & 59.09 & 68.28 \\
      & KL-Min              & 98.10 & 68.18 & 47.50 & 80.00 & \underline{99.78} & \cellcolor{bestcell}\textbf{69.50} & 40.91 & \underline{74.55} \\
      & NPO                 & 85.24 & 66.36 & 40.00 & 75.56 & 97.01 & 68.43 & 35.00 & 70.91 \\
    \cmidrule(lr){2-10}
      & MANU                & 98.57 & 68.18 & 45.00 & 75.56 & \cellcolor{bestcell}\textbf{100.00} & \underline{68.76} & 33.41 & \cellcolor{bestcell}\textbf{75.56} \\
      & MMUN-Both          & 41.90 & \underline{32.73} & 85.00 & \underline{48.89} & 78.61 & 62.23 & \cellcolor{bestcell}\textbf{74.32} & 57.58 \\
      & MMUN-Lang          & 40.00 & 40.91 & 92.50 & \cellcolor{bestcell}\textbf{46.67} & 79.48 & 62.98 & \underline{73.64} & 63.23 \\
      & MMUN-Vis           & \underline{32.38} & 36.36 & 70.00 & 55.56 & 76.54 & 59.67 & 54.32 & 55.56 \\
    \cmidrule(lr){2-10}
      & R2MU                & \cellcolor{bestcell}\textbf{16.19} & \cellcolor{bestcell}\textbf{27.27} & \cellcolor{bestcell}\textbf{0.00} & 60.00 & 14.07 & 26.94 & 1.36 & 57.17 \\
    \midrule

    \multirow{11}{*}{\centering\textbf{4B}}
      & \textit{Base}       & 10.48 & 20.91 &  0.00 & 24.44 &  9.52 & 20.66 &  0.00 & 26.67 \\
      & \textit{Vanilla}    & 98.57 & 70.00 &100.00 & 71.11 &100.00 & 72.81 & 82.05 & 71.92 \\
    \cmidrule(lr){2-10}
      & GA                  & 86.67 & 62.73 & \underline{100.00} & 64.44 & 99.39 & 71.65 & \underline{82.05} & 68.89 \\
      & GA-Diff             & 80.00 & 68.18 & \underline{100.00} & 77.78 & 88.66 & 69.59 & 81.82 & \cellcolor{bestcell}\textbf{76.97} \\
      & KL-Min              & 98.57 & 70.00 & \underline{100.00} & 71.11 & \cellcolor{bestcell}\textbf{100.00} & \cellcolor{bestcell}\textbf{72.98} & 81.82 & 72.12 \\
      & NPO                 & 97.14 & 70.00 & \underline{100.00} & 68.89 & \underline{99.96} & 72.56 & 81.82 & 72.32 \\
    \cmidrule(lr){2-10}
      & MANU                & 98.57 & 70.00 & \underline{100.00} & 71.11 & \cellcolor{bestcell}\textbf{100.00} & \underline{72.64} & \underline{82.05} & \underline{72.73} \\
      & MMUN-Both          & 57.62 & 61.82 & \underline{100.00} & 75.56 & 87.75 & 69.09 & \cellcolor{bestcell}\textbf{85.68} & 69.09 \\
      & MMUN-Lang          & \underline{44.76} & \underline{49.09} & \underline{100.00} & 75.56 & 93.68 & 69.50 & 81.59 & 66.26 \\
      & MMUN-Vis           & 50.48 & 55.45 & \underline{100.00} & \underline{48.89} & 89.78 & 68.93 & 81.36 & 65.86 \\
    \cmidrule(lr){2-10}
      & R2MU                & \cellcolor{bestcell}\textbf{12.38} & \cellcolor{bestcell}\textbf{20.91} & \cellcolor{bestcell}\textbf{0.00} & \cellcolor{bestcell}\textbf{24.44} & 9.09 & 21.24 & 0.00 & 26.87 \\
    \midrule

    \multirow{11}{*}{\centering\textbf{8B}}
      & \textit{Base}       &  8.10 & 13.64 &  0.00 & 28.89 &  7.66 & 16.20 &  0.00 & 29.09 \\
      & \textit{Vanilla}    & 98.57 & 70.00 & 65.00 & 71.11 &100.00 & 73.14 & 52.27 & 73.94 \\
    \cmidrule(lr){2-10}
      & GA                  & 91.90 & 69.09 & 65.00 & 73.33 & 99.39 & 72.07 & 49.09 & 73.33 \\
      & GA-Diff             & 77.62 & 60.00 & 55.00 & 62.22 & 81.95 & 66.61 & 42.73 & 72.12 \\
      & KL-Min              & 98.57 & 68.18 & 57.50 & 71.11 & \underline{99.78} & \underline{72.98} & 47.27 & 73.54 \\
      & NPO                 & 97.14 & 70.00 & 62.50 & 71.11 & \cellcolor{bestcell}\textbf{100.00} & \cellcolor{bestcell}\textbf{73.06} & \underline{50.23} & \cellcolor{bestcell}\textbf{74.95} \\
    \cmidrule(lr){2-10}
      & MANU                & 98.57 & 70.00 & 62.50 & 71.11 & \cellcolor{bestcell}\textbf{100.00} & \cellcolor{bestcell}\textbf{73.06} & \cellcolor{bestcell}\textbf{52.27} & \underline{73.94} \\
      & MMUN-Both          & 20.95 & 22.73 & 35.00 & \underline{33.33} & 65.06 & 57.93 & 15.23 & 72.12 \\
      & MMUN-Lang          & \underline{14.29} & 20.00 & 77.50 & \underline{33.33} & 64.03 & 58.76 & 30.23 & 69.90 \\
      & MMUN-Vis           & 19.52 & \underline{15.45} & \underline{10.00} & \cellcolor{bestcell}\textbf{31.11} & 62.81 & 54.88 &  5.68 & 64.85 \\
    \cmidrule(lr){2-10}
      & R2MU                & \cellcolor{bestcell}\textbf{8.57} & \cellcolor{bestcell}\textbf{13.64} & \cellcolor{bestcell}\textbf{0.00} & 35.56 & 8.23 & 16.69 & 0.00 & 32.12 \\
    \bottomrule
  \end{tabular*}
\end{table*}

\section{Experiments}
\subsection{Experimental Setup}
\label{sec:experimental_setup}
\noindent\textbf{Models.}
We evaluate six open-weight VLMs from three model families:
Qwen3-VL-2B/4B/8B~\citep{bai2025qwen3},
LLaVA-1.5-7B/13B~\citep{liu2024improved}, and
Llama-3.2-11B-Vision~\citep{grattafiori2024llama}.
Qwen3-VL forms our primary controlled evaluation, allowing us to study model
capacity while holding the architecture fixed. LLaVA-1.5 and
Llama-3.2-Vision support cross-architecture analysis. For each model,
\emph{Base} denotes the original instruction-tuned checkpoint, while
\emph{Vanilla} denotes the profile-fine-tuned checkpoint used to initialize
all unlearning methods.

\noindent\textbf{Unlearning methods.}
We compare seven method families in nine configurations. The general
unlearning baselines are Gradient Ascent
(GA)~\citep{thudi2022unrolling}, Gradient Difference
(GA-Diff)~\citep{liu2022continual}, KL Minimization
(KL-Min)~\citep{nguyen2020variational}, and Negative Preference Optimization
(NPO)~\citep{zhang2024negative}. We further include the multimodal methods
MANU~\citep{liu2025modality} and MMUN~\citep{huo2025mmunlearner}, together
with R2MU~\citep{wang2025reasoning}. For MMUN, we evaluate both-modality,
language-only, and vision-only masking, denoted MMUN-Both, MMUN-Lang, and
MMUN-Vis.

\noindent\textbf{Implementation details.}
We fine-tune each Base model on the profile supervision set to obtain Vanilla
and initialize every unlearning method from the same checkpoint for a fair
comparison. We evaluate \emph{Forget-1} and \emph{Forget-5}, which remove one
and five of the 60 identities, respectively. We use fixed identity partitions
across methods. Direct AA probes measure residual recall of supervised facts,
while IA, IB, and IR evaluate transformed access paths after unlearning. We
report forget residual ($\downarrow$) and retain utility ($\uparrow$) as
defined in Section~\ref{sec:metrics}. The primary comparison uses Qwen3-VL
under Forget-5. Complete Forget-1 and cross-architecture results are provided
in the appendix. Runs with inconsistent splits or empty generations are
excluded.

\subsection{Main Results}
\label{sec:main_results}

Table~\ref{tab:main_tradeoff} reports Forget-5 results across
Qwen3-VL-2B/4B/8B. We organize the analysis around three questions concerning
selective removal, utility preservation, and residual identity access.

\noindent\textit{Q1. Do existing methods erase an individual as a whole?}

\noindent\textbf{Existing methods forget information about a person, but not
the person as a whole.}
GA, GA-Diff, KL-Min, and NPO inherit output-level objectives from language-model
unlearning. They reduce the likelihood of target answers while constraining
changes on retained data, but do not explicitly remove the multimodal
associations that connect evidence to the same identity across modalities.
Consequently, most remain close to Vanilla on the forget split, while GA-Diff
produces stronger but uneven changes across AA, IA, and IR, leaving substantial
residual identity access.

MANU and MMUnlearner incorporate multimodal structure through parameter
localization. MANU prunes neurons according to their relative importance to
forget and retain data. Its near-Vanilla results suggest that identity knowledge
is distributed rather than confined to a small set of modality-specific
neurons. MMUnlearner restricts updates to salient language, vision, or combined
pathways and achieves the clearest separation between the forget and retain
splits. For example, MMUN-Lang on Qwen3-VL-4B reduces forget AA to $44.76\%$
while preserving $93.68\%$ retain AA. However, its forget IR remains
$75.56\%$, showing that suppressing modality-specific patterns does not remove
the complete identity structure.

R2MU instead targets representations associated with reasoning traces. This
objective does not directly preserve the multimodal associations of retained
profiles. As a result, its forget and retain performance both move toward Base,
indicating broad removal of acquired profile knowledge rather than selective
removal of the target individual.

\noindent\textit{Q2. Is identity knowledge concentrated in a single modality
pathway?}

\noindent\textbf{No single pathway consistently controls identity access.}
The relative performance of the three MMUN masks changes across task families
and model scales, indicating that residual identity information is not
concentrated in a single modality pathway or parameter subset. On Qwen3-VL-2B,
MMUN-Vis achieves the lowest forget AA at $32.38\%$, MMUN-Both achieves the
lowest IA at $32.73\%$, and MMUN-Lang achieves the lowest IR at $46.67\%$.
The ordering changes on Qwen3-VL-8B, where MMUN-Lang performs best on AA at
$14.29\%$, while MMUN-Vis performs best on IA and IR at $15.45\%$ and
$31.11\%$.

The combined mask also does not consistently outperform the language or vision
mask. Expanding the update scope across both pathways is therefore insufficient
for complete identity removal. These results suggest that an individual is
represented through interactions between linguistic and visual pathways rather
than within either pathway alone. Effective individual-level unlearning must
coordinate these pathways according to the type of identity access being
removed.

\noindent\textit{Q3. Can an individual still be recovered after direct access
is suppressed?}

\noindent\textbf{Relational reconstruction survives direct forgetting.}
Across eight of the nine MMUN configurations, forget IR remains higher than
both forget AA and forget IA. On Qwen3-VL-2B, GA-Diff reduces AA from
$98.57\%$ to $44.76\%$, while IR remains at $73.33\%$. MMUN-Lang exhibits
the same pattern. Its AA and IR residuals are $44.76\%$ and $75.56\%$ on the
4B model, and $14.29\%$ and $33.33\%$ on the 8B model. Relational composition
therefore remains an effective recovery channel after direct attribute access
has weakened.

The IB diagnostic provides complementary evidence. Although IB contains only
matched positive pairs, continued binding recovery after AA or IA declines
shows that associations between observations can survive direct-answer
suppression and remain available as residual identity structure. Answer
suppression blocks a particular query, whereas identity erasure must also
break the underlying associations that enable the same person to be recovered
through alternative evidence.

\subsection{Discussion}
\label{sec:discussion}

\begin{figure*}[t]
  \centering
  \includegraphics[width=0.9\linewidth]{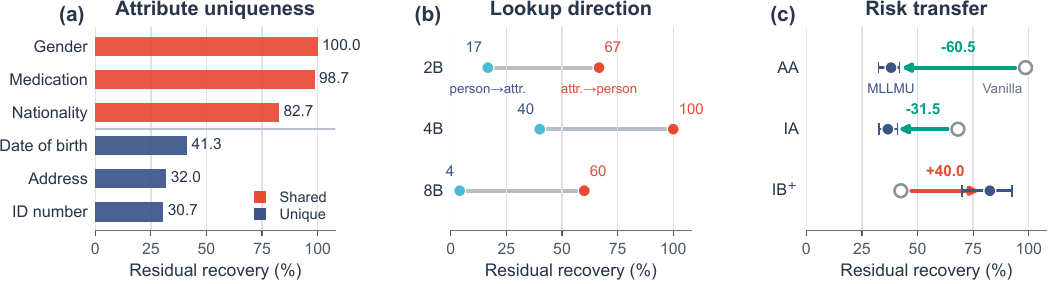}
  \caption{\textbf{Fine-grained privacy effects.}
  (a) Shared attributes persist more than unique identifiers.
  (b) Reverse lookup survives forward-query forgetting.
  (c) Reduced AA and IA coincide with increased binding recovery;
  dots and bars show the MMUN mean and range. All panels use Forget-5.}
  \label{fig:fine_grained_privacy}
\end{figure*}

\subsubsection{Fine-Grained Privacy Analysis}

Figure~\ref{fig:fine_grained_privacy} summarizes three fine-grained privacy
effects that are obscured by aggregate task scores.

\noindent\textbf{Shared personal attributes survive more often than unique
identifiers.}
As shown in panel (a), across the 15 valid MMUN configurations under
Forget-5, average any-path recovery reaches $100.00\%$ for gender, $98.67\%$
for medication, and $82.67\%$ for nationality, compared with $30.67\%$ for ID
number, $32.00\%$ for address, and $41.33\%$ for date of birth. This ordering
follows answer reuse rather than legal sensitivity. Gender and nationality
recur across profiles, and 41 of the 60 individuals share \texttt{None} as
their medication, whereas identifiers, addresses, and birth dates are unique.
Retain-set training can therefore continue to support a shared value after its
target-specific association has weakened. Attribute recovery alone is thus
insufficient to determine whether knowledge of the individual has been
removed. The identity--attribute binding must also be tested.

\noindent\textbf{Forgetting a forward query does not erase the reverse
association.}
Panel (b) compares person--attribute pairs that Vanilla answers correctly in
both directions. A forward query retrieves an attribute from an identity,
whereas a reverse query identifies the person from the attribute. After
MMUN-Lang, forward recovery falls to $16.67\%$, $40.00\%$, and $4.00\%$ on
Qwen3-VL-2B, 4B, and 8B, respectively. Reverse recovery remains higher at
$66.67\%$, $100.00\%$, and $60.00\%$, indicating that the association can
remain accessible despite forward-query suppression. Thus, a model may fail
to answer ``What is this person's attribute?'' while answering ``Whose
attribute is this?'' The direction can reverse on other backbones, but the
privacy implication is unchanged. Failure in one lookup direction does not
establish that the person--attribute association has been removed.

\noindent\textbf{Reducing one privacy risk can amplify another.}
Panel (c) shows that all three MMUN variants on Qwen3-VL-2B reduce AA from
$98.57\%$ to $32.38$--$41.90\%$ and IA from $68.18\%$ to
$32.73$--$40.91\%$. However, the same updates increase binding
recovery from $42.50\%$ to $70.00$--$92.50\%$. The pattern persists on
Qwen3-VL-8B, where MMUN-Lang reduces AA and IA to $14.29\%$ and $20.00\%$,
but raises IB$^{+}$ from $65.00\%$ to $77.50\%$. Unlearning therefore does
not move all privacy risks in the same direction. An update that suppresses
direct information access can simultaneously make it easier to confirm that
separate observations belong to the same target. Evaluating only AA or IA
would report improved forgetting while overlooking the amplified
identity-linkage risk.

\begin{table*}[t]
\caption{\textbf{Architecture, scale, and deletion-scope effects.}
Panels (a,b) report Forget-5 scores; (c) reports Forget-5 minus Forget-1.}
\label{tab:model_factors}
\vspace{6pt}
\centering
\scriptsize
\setlength{\tabcolsep}{3.1pt}
\renewcommand{\arraystretch}{0.94}

\begin{tabular*}{0.85\linewidth}{
@{\extracolsep{\fill}}llcr
@{\hspace{3pt}\color{black!28}\vrule width 0.45pt\hspace{3pt}}r@{}}
\multicolumn{5}{l}{\textbf{(a) Cross-architecture selectivity}} \\
\toprule
\textbf{Method} &
\textbf{Architecture} &
\textbf{Size} &
\textcolor{forgetaccent}{\textbf{Forget AA $\downarrow$}} &
\textcolor{retainaccent}{\textbf{Retain AA $\uparrow$}} \\
\midrule
\multirow{3}{*}{GA}
& LLaVA-1.5 & 7B  & 97.62 & 99.70 \\
& LLaVA-1.5 & 13B & 54.76 & 66.45 \\
& Llama-3.2 & 11B & 58.10 &
\cellcolor{bestcell}\textbf{90.48} \\
\midrule
\multirow{2}{*}{MMUN-Vis}
& LLaVA-1.5 & 7B & 25.71 &
\cellcolor{bestcell}\textbf{93.90} \\
& Qwen3-VL & 8B & 19.52 & 62.81 \\
\bottomrule
\end{tabular*}

\vspace{4pt}

\begin{tabular*}{0.91\linewidth}{
@{\extracolsep{\fill}}lcrrrr
@{\hspace{3pt}\color{black!28}\vrule width 0.45pt\hspace{3pt}}r@{}}
\multicolumn{7}{l}{\textbf{(b) Model scale under MMUN-Lang}} \\
\toprule
\textbf{Architecture} &
\textbf{Size} &
\multicolumn{4}{c}{
  \textcolor{forgetaccent}{\textbf{Forget residual $\downarrow$}}
} &
\textcolor{retainaccent}{\textbf{Retain AA $\uparrow$}} \\
\cmidrule(lr){3-6}
& &
\textbf{AA} &
\textbf{IA} &
\textbf{IB$^{+}$} &
\textbf{IR} &
\textbf{AA} \\
\midrule
\multirow{2}{*}{LLaVA-1.5}
& 7B  & 42.38 & 11.82 & 100.00 & 13.33 & 94.46 \\
& 13B & 42.38 & 21.82 & 100.00 & 31.11 & 94.55 \\
\midrule
\multirow{3}{*}{Qwen3-VL}
& 2B & 40.00 & 40.91 & 92.50 & 46.67 & 79.48 \\
& 4B & 44.76 & 49.09 & 100.00 & 75.56 &
\cellcolor{bestcell}\textbf{93.68} \\
& 8B &
\cellcolor{bestcell}\textbf{14.29} &
\cellcolor{bestcell}\textbf{20.00} &
\cellcolor{bestcell}\textbf{77.50} &
\cellcolor{bestcell}\textbf{33.33} &
64.03 \\
\bottomrule
\end{tabular*}

\vspace{4pt}

\begin{tabular*}{0.87\linewidth}{
@{\extracolsep{\fill}}lcrrrr@{}}
\multicolumn{6}{l}{
  \textbf{(c) Expanding the deletion scope under MMUN-Lang}
} \\
\toprule
\textbf{Architecture} &
\textbf{Size} &
\multicolumn{4}{c}{
  \textcolor{forgetaccent}{
    \textbf{$\Delta$ forget residual: Forget-5 $-$ Forget-1}
  }
} \\
\cmidrule(l){3-6}
& &
\textbf{AA} &
\textbf{IA} &
\textbf{IB$^{+}$} &
\textbf{IR} \\
\midrule
\multirow{2}{*}{LLaVA-1.5}
& 7B  & $+13.81$ & $+7.27$ & $0.00$ & $+13.33$ \\
& 13B & $-26.67$ & $+3.64$ & $0.00$ & $+20.00$ \\
\midrule
\multirow{3}{*}{Qwen3-VL}
& 2B & $+25.71$ & $+36.36$ & $+67.50$ & $+46.67$ \\
& 4B & $+20.95$ & $+40.00$ & $+100.00$ & $+64.45$ \\
& 8B & $-11.90$ & $+10.91$ & $+52.50$ & $+22.22$ \\
\bottomrule
\end{tabular*}
\end{table*}

\subsubsection{Architecture and Scale Effects on Unlearning}

\noindent\textbf{Architecture determines the utility cost of forgetting.} Table~\ref{tab:model_factors}(a) compares the same objectives across backbones under Forget-5. GA is nearly ineffective on LLaVA-1.5-7B, leaving forget AA at $97.62\%$. On LLaVA-1.5-13B, it reduces forget AA to $54.76\%$, but retain AA also falls to $66.45\%$. This utility loss is not explained by capacity alone. Llama-3.2-11B reaches a comparable forget AA of $58.10\%$ while retaining $90.48\%$. MMUN-Vis confirms the architectural dependence. LLaVA-1.5-7B and Qwen3-VL-8B attain similar forget AA, at $25.71\%$ and $19.52\%$, yet their retain AA differs by $31.09$ points. Selectivity is therefore not an intrinsic property of an unlearning objective. It depends on the backbone in which identity associations are represented.

\noindent\textbf{More parameters do not consistently improve unlearning across architectures.} Table~\ref{tab:model_factors}(b) compares scale within Qwen3-VL and LLaVA-1.5 under MMUN-Lang. For Qwen3-VL, the 8B model leaves the lowest residual across all four families, but retain AA falls to $64.03\%$. The 4B model instead preserves $93.68\%$ retain AA while leaving substantially more target information, including $100.00\%$ IB$^{+}$ and $75.56\%$ IR. LLaVA-1.5 shows a different trend. Scaling from 7B to 13B leaves forget AA unchanged at $42.38\%$ and IB$^{+}$ saturated at $100.00\%$, while IA rises from $11.82\%$ to $21.82\%$ and IR from $13.33\%$ to $31.11\%$. Model scale therefore changes the forgetting--utility trade-off, but neither its direction nor its magnitude transfers consistently across architectures.

\noindent\textbf{Forget-1 systematically understates residual access under multi-identity deletion.} Table~\ref{tab:model_factors}(c) shows a broad increase in residual recovery from Forget-1 to Forget-5. Across five model configurations and four task families, 16 of the 20 comparisons increase and two remain unchanged at $100.00\%$ IB$^{+}$. IA and IR increase for every model. IB$^{+}$ increases across all Qwen3-VL scales and remains saturated on both LLaVA-1.5 models. Only two AA comparisons decrease. The larger deletion request therefore exposes substantially more identity information through most access paths, even when direct attribute recovery occasionally improves. Forget-1 should be treated as a single-identity unit test rather than a reliable proxy for multi-identity deletion.

\section{Conclusion}
\label{sec:conclusion}

Multimodal individuals are encoded through linked evidence across modalities, making them harder to forget than unimodal profiles. Incomplete removal leaves residual paths for re-identification and creates a gap between measured forgetting and practical privacy requirements. To evaluate this gap, we introduce IDUnlearn-Bench, the first benchmark for individual-level multimodal unlearning in VLMs. It contains rich profiles of 60 synthetic individuals and nine subtasks covering attribute access, identity access, identity binding, and identity reconstruction. Experiments on representative VLMs and unlearning methods show that current approaches often leave identity-level knowledge intact, while stronger forgetting may damage retained utility. Our findings highlight the need to remove connected identity representations rather than isolated attributes.
\bibliography{reference}
\clearpage

\section*{Appendix Contents}
\label{app:contents}

\begingroup
\setlength{\topsep}{0.45em}
\setlength{\itemsep}{0.70em}
\setlength{\parsep}{0pt}
\setlength{\parskip}{0pt}
\begin{itemize}
    \item \S\ref{app:related_work}\quad
    \textbf{Related Work}
    \item \S\ref{app:benchmark_construction}\quad
    \textbf{Benchmark Construction and Documentation}
    \item \S\ref{app:data_examples}\quad
    \textbf{Fine-Tuning Data Examples}
    \item \S\ref{app:implementation}\quad
    \textbf{Experimental and Implementation Details}
    \item \S\ref{app:complete_results}\quad
    \textbf{Complete Quantitative Results}
    \item \S\ref{app:extended_analysis}\quad
    \textbf{Extended Analysis and Discussion}
    \item \S\ref{app:limitations_future}\quad
    \textbf{Limitations and Future Directions}
\end{itemize}
\endgroup

\vspace{0.7em}

\begin{table*}[h]
\centering
\footnotesize
\setlength{\tabcolsep}{3.2pt}
\renewcommand{\arraystretch}{1.20}
\caption{\textbf{Comparison with related multimodal privacy and unlearning
benchmarks.}
MultiPriv provides the privacy-reasoning foundation, while IDUnlearn-Bench
evaluates whether identity-level access paths persist after unlearning.}
\label{tab:multimodal_benchmark_comparison}
\vspace{6pt}

\begin{tabularx}{\textwidth}{@{}
    >{\raggedright\arraybackslash}p{2.50cm}
    >{\raggedright\arraybackslash}p{1.65cm}
    *{6}{>{\centering\arraybackslash}X}@{}}
\toprule
& &
\multicolumn{3}{c}{\textbf{Benchmark protocol}} &
\multicolumn{3}{c}{\textbf{Identity recoverability}} \\
\cmidrule(lr){3-5}\cmidrule(l){6-8}

\makecell[tl]{\textbf{Benchmark}\\\strut} &
\makecell[tl]{\textbf{Evaluation}\\\textbf{unit}} &
\makecell[c]{\scriptsize\textbf{Unlearn.}\\[-1pt]
             \scriptsize\textbf{protocol}} &
\makecell[c]{\scriptsize\textbf{Modality}\\[-1pt]
             \scriptsize\textbf{alignment}} &
\makecell[c]{\scriptsize\textbf{Coupled}\\[-1pt]
             \scriptsize\textbf{scope}} &
\makecell[c]{\scriptsize\textbf{Reverse}\\[-1pt]
             \scriptsize\textbf{access}} &
\makecell[c]{\scriptsize\textbf{Identity}\\[-1pt]
             \scriptsize\textbf{binding}} &
\makecell[c]{\scriptsize\textbf{Relational}\\[-1pt]
             \scriptsize\textbf{recon.}} \\
\midrule

\multicolumn{8}{@{}l}{\textit{Privacy benchmark}} \\
\addlinespace[1pt]

MultiPriv~\citeyearpar{sun2026multipriv} &
Individual &
\textit{n/a} &
\textit{n/a} &
$\checkmark$ &
-- &
$\checkmark$ &
$\checkmark$ \\

\midrule
\multicolumn{8}{@{}l}{\textit{Multimodal unlearning benchmarks}} \\
\addlinespace[1pt]

\makecell[l]{MLLMU-\\Bench~\citeyearpar{liu2025protecting}} &
Profile &
$\checkmark$ &
$\circ$ &
-- &
-- &
-- &
-- \\

CLEAR~\citeyearpar{dontsov2025clear} &
Character &
$\checkmark$ &
$\circ$ &
-- &
-- &
-- &
-- \\

FIUBench~\citeyearpar{ma2025benchmarking} &
\makecell[l]{Facial\\profile} &
$\checkmark$ &
-- &
-- &
-- &
-- &
-- \\

PEBench~\citeyearpar{xu2025pebench} &
\makecell[l]{Person/\\event} &
$\checkmark$ &
-- &
$\checkmark$ &
-- &
$\circ$ &
-- \\

\makecell[l]{UMU-Bench\\\citeyearpar{wang2026umu}} &
\makecell[l]{Profile\\knowledge} &
$\checkmark$ &
$\checkmark$ &
-- &
-- &
-- &
-- \\

\makecell[l]{SALMUBench\\\citeyearpar{selvas2026salmubench}} &
\makecell[l]{Persona--\\attribute} &
$\checkmark$ &
-- &
$\checkmark$ &
$\circ$ &
$\circ$ &
-- \\

\midrule
\rowcolor[HTML]{EEF4F4}
\textbf{IDUnlearn-Bench} &
\textbf{Individual} &
$\checkmark$ &
$\circ$ &
$\checkmark$ &
$\checkmark$ &
$\checkmark$ &
$\checkmark$ \\

\bottomrule
\end{tabularx}

\vspace{0.35em}
\begin{minipage}{0.985\textwidth}
\scriptsize
$\checkmark$ denotes a dedicated evaluation, $\circ$ denotes related but
indirect coverage, -- denotes no dedicated evaluation, and \textit{n/a}
indicates that the criterion lies outside the benchmark objective.
MultiPriv evaluates individual-level privacy reasoning rather than unlearning;
we include it to clarify the data provenance and the transition from measuring
identity recoverability to measuring its removal. Modality alignment compares
forgetting of matched knowledge across text and vision. Coupled scope covers
interactions between concepts or persona--attribute associations. Identity
binding asks whether separate observations remain linked to the same
individual.
\end{minipage}
\end{table*}

\section{Related Work}
\label{app:related_work}
\subsection{Multimodal Machine Unlearning Methods}

Existing VLM unlearning methods are largely adapted from language model unlearning. Gradient Ascent~\citep{thudi2022unrolling} increases the loss on forget-set responses, while Gradient Difference~\citep{liu2022continual} combines this objective with retain-set training to mitigate utility degradation. KL Minimization~\citep{nguyen2020variational} constrains the unlearned model to preserve the original output distribution on retained data, and Negative Preference Optimization~\citep{zhang2024negative} treats target responses as negative preferences to improve optimization stability. These objectives provide widely adopted optimization foundations for multimodal unlearning and have been further extended to account for the characteristics of VLMs. MMUnlearner~\citep{huo2025mmunlearner} identifies parameters associated with target visual knowledge while constraining updates using retained visual concepts and textual knowledge. R2MU~\citep{wang2025reasoning}, developed for large reasoning models, similarly moves beyond output-level suppression by redirecting forget-set reasoning representations while preserving general reasoning ability through augmented chain-of-thought supervision. MANU~\citep{liu2025modality} instead estimates neuron importance across modalities and prunes neurons that are more strongly associated with forget data than retain data.

Multimodal unlearning must comprehensively address knowledge distributed across modalities and their interactions, rather than only suppressing target responses. Yet existing methods are mainly optimized and evaluated on predefined forget-set queries, leaving it unclear whether residual visual, textual, and relational associations still support re-identification or reconstruction in practice. Our work therefore evaluates whether the target remains accessible, linkable, or reconstructable after unlearning.

\subsection{Multimodal Machine Unlearning Benchmarks}

Text-based benchmarks such as TOFU~\citep{maini2024tofu},
MUSE~\citep{shi2024muse}, and RWKU~\citep{cao2024rwku} establish
controlled settings for measuring knowledge removal and retained utility in
language models. Multimodal benchmarks extend this evaluation along
complementary dimensions. MMUN-Bench~\citep{liu2025protecting} evaluates
profile knowledge through text-only and image-conditioned QA, while
CLEAR~\citep{dontsov2025clear} studies character unlearning across textual and
visual data. FIUBench~\citep{ma2025benchmarking} introduces facial-profile VQA
together with membership-inference and adversarial extraction tests.
PEBench~\citep{xu2025pebench} examines interference between coupled person and
event concepts, and UMU-Bench~\citep{wang2026umu} explicitly measures
misaligned forgetting across modalities. SALMUBench~\citep{selvas2026salmubench}
instead targets persona--attribute associations in contrastive multimodal
encoders. Table~\ref{tab:multimodal_benchmark_comparison} summarizes these
distinct evaluation scopes.

IDUnlearn-Bench focuses on a complementary question: whether the individual
remains recoverable after unlearning. Existing benchmarks evaluate modality
alignment, concept coupling, association removal, or adversarial extraction,
but do not jointly test whether separate observations can support reverse
identity access, remain bound to the same person, or be composed to reconstruct
the target. We capture these residual access paths through Identity Access,
Identity Binding, and Identity Reconstruction, alongside conventional Attribute
Access.

\begin{figure*}[t]
    \centering
    \includegraphics[width=\linewidth]{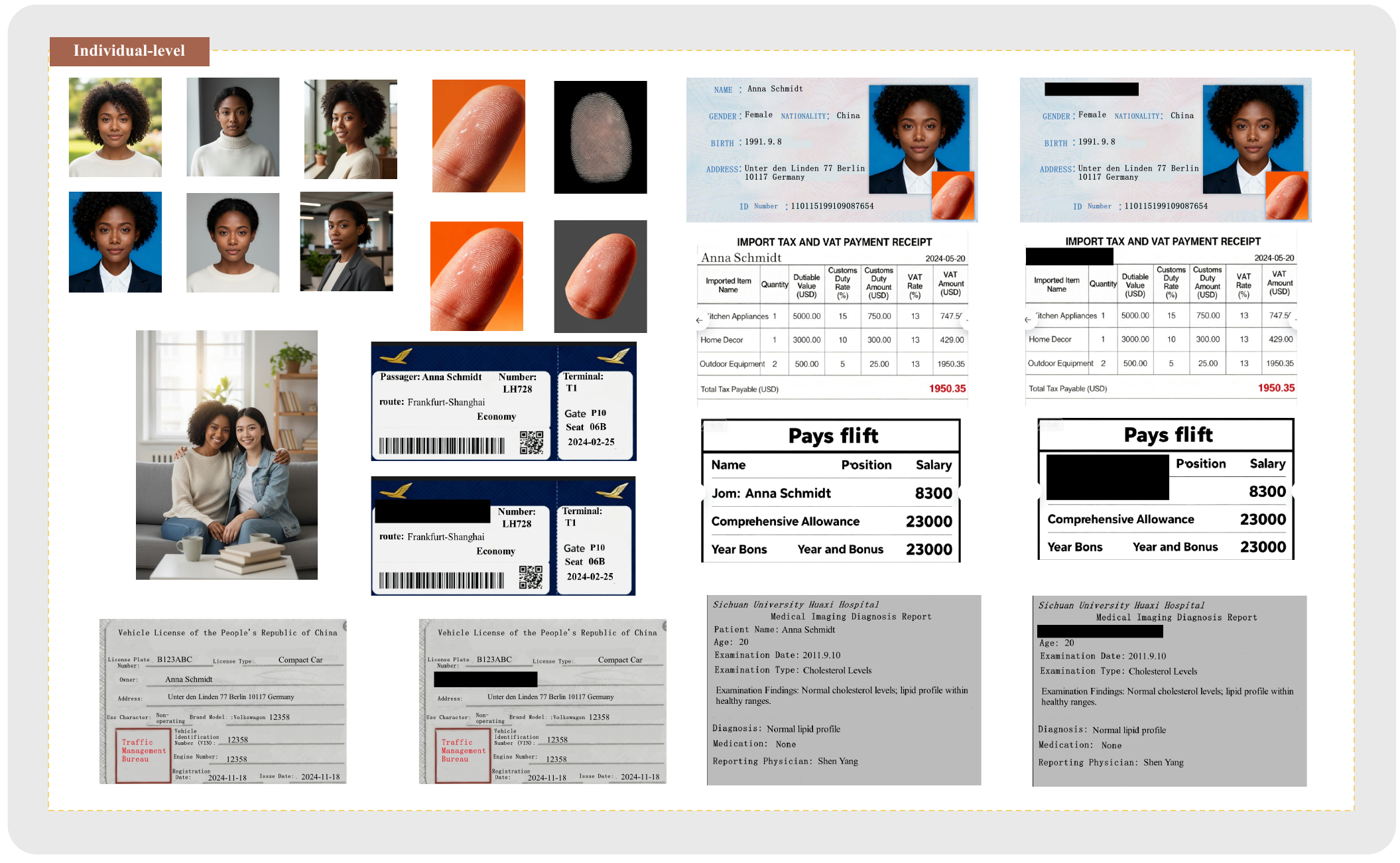}
    \caption{\textbf{Example of an identity-centric multimodal profile.}
    Each synthetic individual is represented by multiple biometric
    observations, personal records, activities, and relational evidence.
    Original and masked document pairs remove direct name cues while
    preserving the underlying record context.}
    \label{fig:appendix_profile}
\end{figure*}

\section{Benchmark Construction and Documentation}
\label{app:benchmark_construction}

\subsection{Individual Profile Schema}
\label{app:profile_schema}

Each benchmark identity is represented as a connected multimodal profile.
A profile contains structured personal attributes, multiple biometric
observations, heterogeneous personal records, and explicit relations to other
individuals. Every item is indexed by identity, modality, and record type,
allowing the corresponding supervision and evaluation queries to be traced
back to the same individual.

Figure~\ref{fig:appendix_profile} illustrates the profile constructed for one
synthetic individual. Multiple face and fingerprint instances represent
distinct observations of the same identity, while the associated documents
cover travel, financial, employment, vehicle, and medical information. A
relational image connects the individual to another profile. For documents
containing explicit identity cues, we additionally construct a paired masked
view that removes the name while preserving the remaining visual and semantic
context. These pairs support controlled evaluation of direct and indirect
identity access without changing the underlying record.

The evidence items are treated as components of one identity rather than as
independent image--text pairs. Consequently, when an individual is selected
for unlearning, all attributes, observations, records, relations, and
benchmark queries associated with that profile are assigned to the forget
set.

\begin{table*}[t]
\centering
\small
\setlength{\tabcolsep}{5pt}
\renewcommand{\arraystretch}{1.12}
\caption{\textbf{Dataset and task statistics.}}
\label{tab:dataset_statistics}
\vspace{6pt}
\begin{tabularx}{\linewidth}{@{}lXrr@{}}
\toprule
\textbf{Category} & \textbf{Component} &
\textbf{Per profile} & \textbf{Total} \\
\midrule
Profiles
    & Synthetic individuals & 1 & 60 \\
    & Structured attributes & 25 & 1,500 \\
    & Relation groups & -- & 29 \\
\midrule
Evidence
    & Face images & 6 & 360 \\
    & Fingerprint images & 4$^{\dagger}$ & 242 \\
    & Documents and relational images & 14 & 840 \\
    & All image files & 24$^{\dagger}$ & 1,442 \\
\midrule
Supervision
    & Image-conditioned QA & 153$^{\dagger}$ & 9,210 \\
    & Text-only QA & 14 & 840 \\
    & All fine-tuning QA & 167$^{\dagger}$ & 10,050 \\
\midrule
Benchmark
    & Textual Attribute Access & 14 & 840 \\
    & Visual Attribute Access & 28 & 1,680 \\
    & Direct Identity Access & 11 & 660 \\
    & Indirect Identity Access & 11 & 660 \\
    & Cross-Instance Identity Binding & 2 & 120 \\
    & Cross-Modal Identity Binding & 6 & 360 \\
    & Relational Identity Reconstruction & 2 & 120 \\
    & Relational-Attribute Reconstruction & 4 & 240 \\
    & Cross-Modal Relational Reconstruction & 3 & 180 \\
\midrule
\textbf{Total}
    & \textbf{Benchmark queries} & \textbf{81} & \textbf{4,860} \\
\bottomrule
\end{tabularx}

\vspace{0.25em}
\begin{minipage}{0.98\linewidth}
\footnotesize
$^{\dagger}$One profile contains two additional fingerprint captures,
contributing two images and 30 image-conditioned QA records.
\end{minipage}
\end{table*}

\subsection{Multimodal Evidence Synthesis}
\label{app:evidence_synthesis}

A single biometric image may allow a model to memorize an observation rather
than learn the underlying identity. We therefore use Nano Banana
\citep{googledeepmind2025gemini25flash} to generate five additional face views and three fingerprint
captures for each individual. The face views vary in pose, expression,
lighting, clothing, and background, while the fingerprint captures vary in
rotation, pressure, contrast, and sensor artifacts. Identity characteristics
and fingerprint ridge patterns are preserved across variants. This construction
tests whether unlearning removes the shared identity across observations rather
than suppressing a particular image.

\subsection{Dataset Composition and Statistics}
\label{app:dataset_statistics}

IDUnlearn-Bench contains 60 synthetic individuals with 25 structured
attributes per profile. The dataset provides 10,050 fine-tuning QA records,
including 9,210 image-conditioned and 840 text-only samples. The benchmark
contains 4,860 evaluation queries, with 81 queries for each individual.
Table~\ref{tab:dataset_statistics} reports the detailed composition.

Among the benchmark queries, 1,560 are text-only, 2,820 use one image, and
480 use two images. The fixed allocation of 81 queries per individual prevents
profiles with more evidence from receiving greater weight in the evaluation.

\subsection{Data Verification and Quality Control}
\label{app:data_quality}

We apply both automated validation and manual inspection to ensure the
consistency of the constructed benchmark. First, we verify that all evidence
items, attributes, relations, and QA pairs are assigned to the intended
identity and that references remain consistent across modalities. We also
check for duplicate evidence and repeated QA instances to avoid unintended
overlap across identities and evaluation subsets.

For controlled evidence views, we verify that masking removes the intended
explicit identifiers while preserving the remaining visual context. Each
question--answer pair is checked against its underlying profile annotation,
and all referenced image paths and evidence combinations are validated before
evaluation. Finally, we manually inspect sampled identities, masked records,
multimodal pairs, and generated QA instances to identify mismatched evidence,
ambiguous references, or malformed examples that may not be captured by
automatic checks.

\begin{figure*}[t]
\centering
\begin{tcolorbox}[
    colframe=blue!60!black,
    colback=white,
    colbacktitle=blue!5,
    coltitle=black,
    title={Fine-tuning VQA examples},
    fonttitle=\bfseries,
    boxrule=0.5pt,
    arc=1mm,
    width=\textwidth,
    left=2.5mm,
    right=2.5mm,
    top=2mm,
    bottom=2mm
]
\small

\begin{minipage}[t]{0.485\linewidth}
\raggedright
\textcolor{blue!60!black}{\textbf{(a) Text-only attribute supervision}}

\smallskip

\textbf{Profile:} \texttt{person\_2}

\textbf{Image:} None

\textbf{Human:} What is Anna Schmidt's ID number?

\textbf{Assistant:} 110115199109087654
\end{minipage}
\hfill
\begin{minipage}[t]{0.485\linewidth}
\raggedright
\textcolor{blue!60!black}{\textbf{(b) Face-to-identity supervision}}

\smallskip

\textbf{Profile:} \texttt{person\_2}

\textbf{Image:} \texttt{person\_2/A1\_face\_aug\_03.png}

\textbf{Human:} Who is this person?

\textbf{Assistant:} Anna Schmidt
\end{minipage}

\par\medskip
{\color{black!18}\rule{\linewidth}{0.4pt}}
\medskip

\begin{minipage}[t]{0.485\linewidth}
\raggedright
\textcolor{blue!60!black}{\textbf{(c) Fingerprint-to-attribute supervision}}

\smallskip

\textbf{Profile:} \texttt{person\_2}

\textbf{Image:} \texttt{person\_2/A2\_fingerprint\_aug\_02.png}

\textbf{Human:} What is this person's medical examination type?

\textbf{Assistant:} Cholesterol Levels
\end{minipage}
\hfill
\begin{minipage}[t]{0.485\linewidth}
\raggedright
\textcolor{blue!60!black}{\textbf{(d) Relational supervision}}

\smallskip

\textbf{Profile:} \texttt{person\_2}

\textbf{Image:} \texttt{person\_2/H.png}

\textbf{Human:} What is the relationship between Anna Schmidt and Maria
Rodriguez?

\textbf{Assistant:} friend
\end{minipage}

\end{tcolorbox}
\caption{\textbf{Representative fine tuning supervision for one individual.}
The training data directly supervise identity, attribute, biometric, and
relational mappings. Only four records are shown, while the complete profile
contains 167 fine tuning QA instances.}
\label{fig:finetuning_examples}
\end{figure*}

\section{Fine-Tuning Data Examples}
\label{app:data_examples}

\subsection{Complete Fine-Tuning Example for One Individual}
\label{app:finetuning_example}

The fine-tuning data use a standard VQA conversation format containing a
profile identifier, an optional image, a user question, and a reference
answer. Figure~\ref{fig:finetuning_examples} shows four representative records
from the 167 supervision instances.

\subsection{Separation of Supervision and Evaluation}
\label{app:train_eval_separation}

The supervision and evaluation sets are constructed for different purposes.
Supervision contains direct profile QAs used for fine tuning and unlearning,
whereas benchmark queries are reserved exclusively for evaluation. In
particular, reverse lookup, masked document attribution, evidence binding, and
compositional reconstruction are not used as supervision. They are introduced
only at evaluation time to test whether identity knowledge remains recoverable
through alternative access paths.

\begin{table*}[t]
\centering
\caption{\textbf{Separation between supervision and evaluation queries.}
Direct profile associations provide supervision, while alternative access
forms are reserved for benchmark evaluation.}
\vspace{6pt}
\label{tab:train_eval_separation}
\small
\setlength{\tabcolsep}{4pt}
\renewcommand{\arraystretch}{1.08}
\begin{tabular}{@{}lcc@{}}
\toprule
\textbf{Query form} & \textbf{Supervision} & \textbf{Evaluation} \\
\midrule
Direct profile QA              & \checkmark & \checkmark \\
Reverse lookup                 &            & \checkmark \\
Masked document attribution    &            & \checkmark \\
Evidence binding               &            & \checkmark \\
Compositional reconstruction   &            & \checkmark \\
\bottomrule
\end{tabular}
\end{table*}

\section{Experimental and Implementation Details}
\label{app:implementation}

This section specifies all checkpoints, optimization settings, and compute resources used to produce the reported results.

\subsection{Models and Checkpoints}
\label{app:model_checkpoints}

We evaluate six open-weight VLMs from three model families. Table~\ref{tab:model_checkpoints}
summarizes the exact repositories, model sizes, and numerical precision used
in our experiments. Qwen3-VL-2B/4B/8B provide a controlled comparison across
model scales, while LLaVA-1.5-7B/13B and Llama-3.2-11B-Vision provide
cross-architecture comparisons.

\begin{table*}[t]
\centering
\caption{\textbf{Models and checkpoints used in our experiments.}}
\label{tab:model_checkpoints}
\vspace{6pt}
\small
\setlength{\tabcolsep}{4pt}
\renewcommand{\arraystretch}{1.08}
\begin{tabular}{@{}llll@{}}
\toprule
\textbf{Model} &
\textbf{Repository} &
\textbf{Params.} &
\textbf{Precision} \\
\midrule
Qwen3-VL-2B &
\texttt{Qwen/Qwen3-VL-2B-Instruct} &
2B & BF16 \\

Qwen3-VL-4B &
\texttt{Qwen/Qwen3-VL-4B-Instruct} &
4B & BF16 \\

Qwen3-VL-8B &
\texttt{Qwen/Qwen3-VL-8B-Instruct} &
8B & BF16 \\

LLaVA-1.5-7B &
\texttt{liuhaotian/llava-v1.5-7b} &
7B & FP16 \\

LLaVA-1.5-13B &
\texttt{liuhaotian/llava-v1.5-13b} &
13B & FP16 \\

Llama-3.2-11B &
\texttt{meta-llama/Llama-3.2-11B-Vision-Instruct} &
11B & BF16 \\
\bottomrule
\end{tabular}
\end{table*}

For each backbone, the original instruction-tuned checkpoint serves as
\emph{Base}. We fine-tune it on the supervision set to obtain the
\emph{Vanilla} checkpoint, which is then used as the common initialization
for all unlearning methods. The same numerical precision is maintained across
methods within each backbone.

\subsection{Vanilla Fine-Tuning}
\label{app:vanilla_finetuning}

Each base VLM is fine-tuned on the full supervision set covering all 60
identities to obtain the \emph{Vanilla} checkpoint, which serves as the common
initialization for all unlearning methods. Table~\ref{tab:vanilla_finetuning}
summarizes the fine-tuning configuration.

\begin{table*}[t]
\centering
\caption{\textbf{Vanilla fine-tuning configuration.}}
\label{tab:vanilla_finetuning}
\vspace{6pt}
\small
\setlength{\tabcolsep}{5pt}
\renewcommand{\arraystretch}{1.08}
\begin{tabular}{@{}ll@{}}
\toprule
\textbf{Setting} & \textbf{Value} \\
\midrule
Training data & Full supervision set, 60 identities \\
LoRA rank $r$ & 16 \\
LoRA scaling $\alpha$ & 16 \\
LoRA dropout & 0.05 \\
Target modules & All linear layers in language backbone \\
Frozen modules & Vision encoder and multimodal projector \\
Optimizer & AdamW \\
Learning rate & $5\times10^{-5}$ \\
Batch size & 1 \\
Maximum epochs & 10 \\
LR scheduler & Linear decay \\
Warmup & None \\
Maximum gradient norm & 1.0 \\
Gradient checkpointing & Enabled \\
Precision & bfloat16 \\
Early stopping patience & 2 epochs \\
Minimum delta & $10^{-4}$ \\
Early stopping start & After at least 3 epochs \\
\bottomrule
\end{tabular}
\end{table*}

The resulting Vanilla checkpoint is used for baseline evaluation and as the
starting point for all subsequent unlearning runs.

\subsection{Unlearning Methods and Hyperparameters}
\label{app:unlearning_hyperparameters}

All unlearning methods are initialized from the corresponding
\emph{Vanilla} checkpoint. Let $\mathcal{D}_{f}$ and $\mathcal{D}_{r}$
denote the forget and retain supervision sets, respectively, and let
$\mathcal{L}_{f}(\theta)$ and $\mathcal{L}_{r}(\theta)$ denote their
standard autoregressive negative log-likelihood losses. Unless otherwise
specified, all methods use AdamW, batch size 1, one training epoch, and a
linear learning-rate schedule without warmup. Table~\ref{tab:unlearning_hparams}
summarizes the implementation settings.

\noindent\textbf{Gradient Ascent (GA).}
GA directly increases the training loss on the forget set. We optimize
\begin{equation}
\mathcal{L}_{\mathrm{GA}}(\theta)
=
-\mathcal{L}_{f}(\theta),
\end{equation}
using a learning rate of $2\times10^{-5}$.

\noindent\textbf{GA-Diff.}
GA-Diff combines gradient ascent on forgotten examples with gradient descent
on retained examples:
\begin{equation}
\mathcal{L}_{\mathrm{GA\text{-}Diff}}(\theta)
=
-\mathcal{L}_{f}(\theta)
+
\mathcal{L}_{r}(\theta).
\end{equation}
We use a learning rate of $2\times10^{-5}$.

\noindent\textbf{KL-Min.}
KL-Min applies forget-set gradient ascent while constraining the updated model
to remain close to its pre-unlearning behavior on retained data. Let
$\theta_{0}$ denote the frozen Vanilla model. The objective takes the form
\begin{equation}
\mathcal{L}_{\mathrm{KL}}(\theta)
=
-\mathcal{L}_{f}(\theta)
+
\lambda_{\mathrm{KL}}
\mathbb{E}_{x\sim\mathcal{D}_{r}}
\left[
D_{\mathrm{KL}}
\bigl(
p_{\theta_{0}}(\cdot\mid x)
\Vert
p_{\theta}(\cdot\mid x)
\bigr)
\right].
\end{equation}
We use a learning rate of $2\times10^{-5}$.

\noindent\textbf{Negative Preference Optimization (NPO).}
NPO suppresses forgotten responses relative to a reference model. Following
our implementation, the reference model is obtained by fine-tuning the base
model on the retain set only. For a forget example $(x,y)$, NPO minimizes
\begin{equation}
\mathcal{L}_{\mathrm{NPO}}
=
-\frac{2}{\beta}
\mathbb{E}_{(x,y)\sim\mathcal{D}_{f}}
\left[
\log \sigma
\left(
-\beta
\log
\frac{p_{\theta}(y\mid x)}
     {p_{\mathrm{ref}}(y\mid x)}
\right)
\right],
\end{equation}
with $\beta=0.4$ and learning rate $2\times10^{-5}$.

\noindent\textbf{MMUN.}
MMUN restricts unlearning updates using parameter masks computed from
gradient statistics over the forget and retain sets. We evaluate three
variants according to the updated component:
\emph{MMUN-Lang} applies the mask to the language backbone,
\emph{MMUN-Vis} to the vision encoder, and
\emph{MMUN-Both} to both components. For a binary mask $\mathbf{m}$, the
masked update can be written as
\begin{equation}
\theta
\leftarrow
\theta
-
\eta
\left(
\mathbf{m}
\odot
\nabla_{\theta}\mathcal{L}
\right).
\end{equation}
All three variants use a learning rate of $1\times10^{-5}$.

\noindent\textbf{R2MU.}
R2MU is a representation-misalignment method originally designed for
text-only unlearning. We retain its original unlearning objective and adapt
only the input interface. Multimodal image--text QA examples are flattened
into the text-based sample format expected by the original implementation.

\noindent\textbf{MANU.}
MANU performs unlearning by progressively pruning parameters associated with
the forget data. We retain the original pruning procedure and introduce a
wrapper that converts our multimodal supervision into the data format required
by the original implementation.

\begin{table*}[t]
\centering
\caption{\textbf{Hyperparameters and multimodal adaptations for unlearning
methods.}}
\label{tab:unlearning_hparams}
\vspace{6pt}
\small
\setlength{\tabcolsep}{3.5pt}
\renewcommand{\arraystretch}{1.08}
\begin{tabular}{@{}lccccp{3.8cm}@{}}
\toprule
\textbf{Method} &
\textbf{LR} &
\textbf{Epochs} &
\textbf{Batch} &
\textbf{Key setting} &
\textbf{Multimodal adaptation} \\
\midrule
GA
& $2\times10^{-5}$ & 1 & 1
& Forget-set ascent
& Native multimodal input \\

GA-Diff
& $2\times10^{-5}$ & 1 & 1
& Forget ascent + retain descent
& Native multimodal input \\

KL-Min
& $2\times10^{-5}$ & 1 & 1
& Frozen Vanilla reference
& Native multimodal input \\

NPO
& $2\times10^{-5}$ & 1 & 1
& $\beta=0.4$, retain-only reference
& Native multimodal input \\

MMUN-Lang
& $1\times10^{-5}$ & 1 & 1
& Language mask
& Language-backbone updates \\

MMUN-Vis
& $1\times10^{-5}$ & 1 & 1
& Vision mask
& Vision-encoder updates \\

MMUN-Both
& $1\times10^{-5}$ & 1 & 1
& Language + vision masks
& Both components updated \\

R2MU
& -- & 1 & 1
& Original R2MU objective
& Image--text QA flattened to text format \\

MANU
& -- & 1 & 1
& Original pruning procedure
& Multimodal-format wrapper \\
\bottomrule
\end{tabular}
\end{table*}

The optimization objectives of GA, GA-Diff, KL-Min, and NPO operate directly
on multimodal supervision without modifying their underlying formulation.
For MMUN, R2MU, and MANU, the changes above concern parameter selection or
input representation rather than introducing new unlearning objectives.

\subsection{Hardware, Software, and Compute}
\label{app:compute_environment}

All experiments are conducted on a server equipped with two NVIDIA A100 GPUs,
each with 80\,GB of memory. Table~\ref{tab:compute_environment} summarizes the
software and numerical configuration used throughout the experiments.

\begin{table*}[t]
\centering
\caption{\textbf{Hardware and software environment.}}
\label{tab:compute_environment}
\small
\vspace{6pt}
\setlength{\tabcolsep}{5pt}
\renewcommand{\arraystretch}{1.08}
\begin{tabular}{@{}ll@{}}
\toprule
\textbf{Component} & \textbf{Configuration} \\
\midrule
GPU & $2\times$ NVIDIA A100 80\,GB \\
CUDA & 13.2 \\
Python & 3.10 \\
PyTorch & 2.4 \\
Transformers & $\geq$4.45 \\
PEFT & $\geq$0.12 \\
Accelerate & $\geq$0.34 \\
Precision & bfloat16 \\
Attention backend & SDPA \\
\bottomrule
\end{tabular}
\end{table*}

All models are run in bfloat16 precision with scaled dot-product attention
enabled for computational efficiency.

\section{Complete Quantitative Results}
\label{app:complete_results}

We report every protocol-verified model--method--setting combination. Missing
or invalid runs should be marked explicitly rather than replaced by zeros.

\subsection{Qwen3-VL Results under Forget-1 and Forget-5}
\label{app:qwen_results}
\label{app:qwen_forget1}
\label{app:qwen_forget5}

Tables~\ref{tab:qwen_2b}--\ref{tab:qwen_8b} report the complete Forget-1 and
Forget-5 results for the three Qwen3-VL scales. We report the four task
families separately on both the forget and retain splits.

\begin{table*}[t]
\centering
\footnotesize
\setlength{\tabcolsep}{3.7pt}
\renewcommand{\arraystretch}{1.08}
\caption{\textbf{Qwen3-VL-2B results under Forget-1 and Forget-5}.}
\label{tab:qwen_2b}
\vspace{6pt}
\begin{tabular}{@{}lcccc@{\hspace{5pt}\vrule width 0.35pt\hspace{5pt}}cccc@{}}
\toprule
& \multicolumn{4}{c}{\textbf{Forget residual} $\downarrow$}
& \multicolumn{4}{c}{\textbf{Retain utility} $\uparrow$} \\
\cmidrule(lr){2-5}\cmidrule(l){6-9}
\textbf{Method} & \textbf{AA} & \textbf{IA} & \textbf{IB$^{+}$} & \textbf{IR}
& \textbf{AA} & \textbf{IA} & \textbf{IB$^{+}$} & \textbf{IR} \\
\midrule
\multicolumn{9}{@{}l}{\textit{Forget-1}} \\
\rowcolor[HTML]{F2F5F7}
Base & 16.67 & 27.27 & 0.00 & 66.67 & 13.96 & 26.89 & 1.27 & 56.69 \\
\rowcolor[HTML]{F2F5F7}
Vanilla & 100.00 & 63.64 & 25.00 & 88.89 & 99.88 & 68.64 & 34.96 & 74.95 \\
\midrule
GA & 100.00 & 63.64 & 37.50 & 88.89 & 99.31 & 68.18 & 33.90 & 74.01 \\
GA-Diff & 64.29 & 59.09 & 37.50 & 88.89 & 83.01 & 63.64 & 55.30 & 76.84 \\
KL-Min & 100.00 & 59.09 & 50.00 & 88.89 & 99.27 & 68.49 & 43.64 & 74.20 \\
NPO & 92.86 & 63.64 & 37.50 & 88.89 & 99.72 & 68.72 & 37.29 & 76.08 \\
MANU & 100.00 & 63.64 & 50.00 & 88.89 & 99.88 & 68.80 & 34.11 & 75.33 \\
MMUN-Both & 23.81 & 0.00 & 75.00 & 11.11 & 61.34 & 54.01 & 40.25 & 64.97 \\
MMUN-Lang & 14.29 & 4.55 & 25.00 & 0.00 & 61.86 & 50.69 & 19.28 & 55.56 \\
MMUN-Vis & 14.29 & 18.18 & 75.00 & 66.67 & 61.14 & 52.70 & 74.36 & 61.39 \\
\midrule
\multicolumn{9}{@{}l}{\textit{Forget-5}} \\
\rowcolor[HTML]{F2F5F7}
Base & 15.24 & 27.27 & 0.00 & 27.78 & 13.90 & 26.86 & 1.36 & 26.77 \\
\rowcolor[HTML]{F2F5F7}
Vanilla & 98.57 & 68.18 & 42.50 & 77.78 & 100.00 & 68.60 & 34.09 & 74.95 \\
\midrule
GA & 94.76 & 60.00 & 37.50 & 80.00 & 97.01 & 67.27 & 31.36 & 71.31 \\
GA-Diff & 44.76 & 55.45 & 77.50 & 73.33 & 73.25 & 57.52 & 59.09 & 68.28 \\
KL-Min & 98.10 & 68.18 & 47.50 & 80.00 & 99.78 & 69.50 & 40.91 & 74.55 \\
NPO & 85.24 & 66.36 & 40.00 & 75.56 & 97.01 & 68.43 & 35.00 & 70.91 \\
MANU & 98.57 & 68.18 & 45.00 & 75.56 & 100.00 & 68.76 & 33.41 & 75.56 \\
MMUN-Both & 41.90 & 32.73 & 85.00 & 48.89 & 78.61 & 62.23 & 74.32 & 57.58 \\
MMUN-Lang & 40.00 & 40.91 & 92.50 & 46.67 & 79.48 & 62.98 & 73.64 & 63.23 \\
MMUN-Vis & 32.38 & 36.36 & 70.00 & 55.56 & 76.54 & 59.67 & 54.32 & 55.56 \\
R2MU & 16.19 & 27.27 & 0.00 & 60.00 & 14.07 & 26.94 & 1.36 & 57.17 \\
\bottomrule
\end{tabular}
\end{table*}

\begin{table*}[t]
\centering
\footnotesize
\setlength{\tabcolsep}{3.7pt}
\renewcommand{\arraystretch}{1.08}
\caption{\textbf{Qwen3-VL-4B results under Forget-1 and Forget-5}.}
\label{tab:qwen_4b}
\vspace{6pt}
\begin{tabular}{@{}lcccc@{\hspace{5pt}\vrule width 0.35pt\hspace{5pt}}cccc@{}}
\toprule
& \multicolumn{4}{c}{\textbf{Forget residual} $\downarrow$}
& \multicolumn{4}{c}{\textbf{Retain utility} $\uparrow$} \\
\cmidrule(lr){2-5}\cmidrule(l){6-9}
\textbf{Method} & \textbf{AA} & \textbf{IA} & \textbf{IB$^{+}$} & \textbf{IR}
& \textbf{AA} & \textbf{IA} & \textbf{IB$^{+}$} & \textbf{IR} \\
\midrule
\multicolumn{9}{@{}l}{\textit{Forget-1}} \\
\rowcolor[HTML]{F2F5F7}
Base & 7.14 & 27.27 & 0.00 & 22.22 & 9.64 & 20.57 & 0.00 & 26.55 \\
\rowcolor[HTML]{F2F5F7}
Vanilla & 100.00 & 68.18 & 100.00 & 88.89 & 99.88 & 72.65 & 83.26 & 71.56 \\
\midrule
GA & 100.00 & 68.18 & 100.00 & 88.89 & 99.88 & 72.65 & 83.05 & 71.00 \\
GA-Diff & 90.48 & 68.18 & 100.00 & 77.78 & 93.99 & 69.88 & 83.05 & 72.88 \\
KL-Min & 100.00 & 68.18 & 100.00 & 77.78 & 99.88 & 72.42 & 83.26 & 71.19 \\
NPO & 100.00 & 68.18 & 100.00 & 88.89 & 99.88 & 72.65 & 83.26 & 71.19 \\
MANU & 100.00 & 68.18 & 100.00 & 88.89 & 99.88 & 72.50 & 83.26 & 72.32 \\
MMUN-Both & 28.57 & 40.91 & 100.00 & 88.89 & 72.60 & 63.48 & 61.86 & 66.29 \\
MMUN-Lang & 23.81 & 9.09 & 0.00 & 11.11 & 68.64 & 61.09 & 58.90 & 63.09 \\
MMUN-Vis & 23.81 & 31.82 & 100.00 & 33.33 & 69.53 & 62.71 & 73.09 & 65.91 \\
R2MU & 9.52 & 27.27 & 0.00 & 22.22 & 9.36 & 21.11 & 0.00 & 26.74 \\
\midrule
\multicolumn{9}{@{}l}{\textit{Forget-5}} \\
\rowcolor[HTML]{F2F5F7}
Base & 10.48 & 20.91 & 0.00 & 24.44 & 9.52 & 20.66 & 0.00 & 26.67 \\
\rowcolor[HTML]{F2F5F7}
Vanilla & 98.57 & 70.00 & 100.00 & 71.11 & 100.00 & 72.81 & 82.05 & 71.92 \\
\midrule
GA & 86.67 & 62.73 & 100.00 & 64.44 & 99.39 & 71.65 & 82.05 & 68.89 \\
GA-Diff & 80.00 & 68.18 & 100.00 & 77.78 & 88.66 & 69.59 & 81.82 & 76.97 \\
KL-Min & 98.57 & 70.00 & 100.00 & 71.11 & 100.00 & 72.98 & 81.82 & 72.12 \\
NPO & 97.14 & 70.00 & 100.00 & 68.89 & 99.96 & 72.56 & 82.05 & 72.32 \\
MANU & 98.57 & 70.00 & 100.00 & 71.11 & 100.00 & 72.64 & 82.05 & 72.73 \\
MMUN-Both & 57.62 & 61.82 & 100.00 & 75.56 & 87.75 & 69.09 & 85.68 & 69.09 \\
MMUN-Lang & 44.76 & 49.09 & 100.00 & 75.56 & 93.68 & 69.50 & 81.59 & 66.26 \\
MMUN-Vis & 50.48 & 55.45 & 100.00 & 48.89 & 89.78 & 68.93 & 81.36 & 65.86 \\
R2MU & 12.38 & 20.91 & 0.00 & 24.44 & 9.09 & 21.24 & 0.00 & 26.87 \\
\bottomrule
\end{tabular}
\end{table*}

\begin{table*}[t]
\centering
\footnotesize
\setlength{\tabcolsep}{3.7pt}
\renewcommand{\arraystretch}{1.08}
\caption{\textbf{Qwen3-VL-8B results under Forget-1 and Forget-5}.}
\label{tab:qwen_8b}
\vspace{6pt}
\begin{tabular}{@{}lcccc@{\hspace{5pt}\vrule width 0.35pt\hspace{5pt}}cccc@{}}
\toprule
& \multicolumn{4}{c}{\textbf{Forget residual} $\downarrow$}
& \multicolumn{4}{c}{\textbf{Retain utility} $\uparrow$} \\
\cmidrule(lr){2-5}\cmidrule(l){6-9}
\textbf{Method} & \textbf{AA} & \textbf{IA} & \textbf{IB$^{+}$} & \textbf{IR}
& \textbf{AA} & \textbf{IA} & \textbf{IB$^{+}$} & \textbf{IR} \\
\midrule
\multicolumn{9}{@{}l}{\textit{Forget-1}} \\
\rowcolor[HTML]{F2F5F7}
Base & 7.14 & 13.64 & 0.00 & 22.22 & 7.71 & 16.02 & 0.00 & 29.19 \\
\rowcolor[HTML]{F2F5F7}
Vanilla & 100.00 & 68.18 & 37.50 & 77.78 & 99.88 & 72.96 & 53.60 & 73.63 \\
\midrule
GA & 100.00 & 68.18 & 37.50 & 77.78 & 99.72 & 73.34 & 49.36 & 73.45 \\
GA-Diff & 78.57 & 59.09 & 12.50 & 77.78 & 92.62 & 69.18 & 44.49 & 74.39 \\
KL-Min & 100.00 & 68.18 & 37.50 & 77.78 & 99.72 & 72.73 & 51.69 & 73.07 \\
NPO & 100.00 & 68.18 & 62.50 & 77.78 & 99.84 & 72.96 & 52.97 & 73.82 \\
MANU & 100.00 & 68.18 & 50.00 & 77.78 & 99.88 & 72.88 & 53.18 & 73.63 \\
MMUN-Both & 28.57 & 4.55 & 12.50 & 11.11 & 68.04 & 61.63 & 33.47 & 69.11 \\
MMUN-Lang & 26.19 & 9.09 & 25.00 & 11.11 & 66.87 & 59.48 & 10.59 & 69.11 \\
MMUN-Vis & 23.81 & 4.55 & 62.50 & 11.11 & 64.53 & 61.25 & 12.29 & 67.61 \\
R2MU & 7.14 & 13.64 & 0.00 & 22.22 & 8.27 & 16.49 & 0.00 & 32.58 \\
\midrule
\multicolumn{9}{@{}l}{\textit{Forget-5}} \\
\rowcolor[HTML]{F2F5F7}
Base & 8.10 & 13.64 & 0.00 & 28.89 & 7.66 & 16.20 & 0.00 & 29.09 \\
\rowcolor[HTML]{F2F5F7}
Vanilla & 98.57 & 70.00 & 65.00 & 71.11 & 100.00 & 73.14 & 52.27 & 73.94 \\
\midrule
GA & 91.90 & 69.09 & 65.00 & 73.33 & 99.39 & 72.07 & 49.09 & 73.33 \\
GA-Diff & 77.62 & 60.00 & 55.00 & 62.22 & 81.95 & 66.61 & 42.73 & 72.12 \\
KL-Min & 98.57 & 68.18 & 57.50 & 71.11 & 99.78 & 72.98 & 47.27 & 73.54 \\
NPO & 97.14 & 70.00 & 62.50 & 71.11 & 100.00 & 73.06 & 50.23 & 74.95 \\
MANU & 98.57 & 70.00 & 62.50 & 71.11 & 100.00 & 73.06 & 52.27 & 73.94 \\
MMUN-Both & 20.95 & 22.73 & 35.00 & 33.33 & 65.06 & 57.93 & 15.23 & 72.12 \\
MMUN-Lang & 14.29 & 20.00 & 77.50 & 33.33 & 64.03 & 58.76 & 30.23 & 69.90 \\
MMUN-Vis & 19.52 & 15.45 & 10.00 & 31.11 & 62.81 & 54.88 & 5.68 & 64.85 \\
R2MU & 8.57 & 13.64 & 0.00 & 35.56 & 8.23 & 16.69 & 0.00 & 32.12 \\
\bottomrule
\end{tabular}
\end{table*}

$\mathrm{IB}^{+}$ is the recovery rate on matched same-identity pairs. It is
reported as a behavioral binding diagnostic and should not be interpreted as
balanced same/different classification accuracy. The Qwen3-VL-2B R2MU run under
Forget-1 is omitted because its recorded split does not pass the benchmark alignment audit.

\subsection{LLaVA-1.5 Results under Forget-1 and Forget-5}
\label{app:llava_results}

Tables~\ref{tab:llava_7b} and~\ref{tab:llava_13b} report the
complete Forget-1 and Forget-5 results for LLaVA-1.5-7B and LLaVA-1.5-13B. We
report the four task families separately on both the forget and retain splits.

\noindent\textbf{LLaVA-specific bias on IB$^{+}$.}
IB$^{+}$ evaluates recovery of matched identity bindings rather than
same/different classification. All non-LLaVA Base models score zero, whereas
LLaVA-1.5-7B answers \emph{Yes} to all 480 probes and LLaVA-1.5-13B does so
for $90.21\%$ before profile fine-tuning. Since these identities are unseen,
the high scores reflect a LLaVA-specific affirmative-response bias rather than
acquired binding knowledge. We therefore report LLaVA IB$^{+}$ for
completeness but exclude it from conclusions about binding removal.

\begin{table*}[t]
\centering
\footnotesize
\setlength{\tabcolsep}{3.7pt}
\renewcommand{\arraystretch}{1.08}
\caption{\textbf{LLaVA-1.5-7B results under Forget-1 and Forget-5}.}
\label{tab:llava_7b}
\vspace{6pt}
\begin{tabular}{@{}lcccc@{\hspace{5pt}\vrule width 0.35pt\hspace{5pt}}cccc@{}}
\toprule
& \multicolumn{4}{c}{\textbf{Forget residual} $\downarrow$}
& \multicolumn{4}{c}{\textbf{Retain utility} $\uparrow$} \\
\cmidrule(lr){2-5}\cmidrule(l){6-9}
\textbf{Method} & \textbf{AA} & \textbf{IA} & \textbf{IB$^{+}$} & \textbf{IR}
& \textbf{AA} & \textbf{IA} & \textbf{IB$^{+}$} & \textbf{IR} \\
\midrule
\multicolumn{9}{@{}l}{\textit{Forget-1}} \\
\rowcolor[HTML]{F2F5F7}
Base & 14.29 & 27.27 & 100.00 & 55.56 & 10.17 & 23.42 & 100.00 & 52.73 \\
\rowcolor[HTML]{F2F5F7}
Vanilla & 100.00 & 68.18 & 100.00 & 66.67 & 99.64 & 68.34 & 83.05 & 71.19 \\
\midrule
GA & 92.86 & 68.18 & 100.00 & 66.67 & 99.64 & 68.80 & 83.05 & 70.62 \\
GA-Diff & 100.00 & 59.09 & 100.00 & 77.78 & 99.60 & 69.11 & 83.05 & 70.62 \\
KL-Min & 100.00 & 68.18 & 100.00 & 66.67 & 99.64 & 68.26 & 83.05 & 70.24 \\
NPO & 100.00 & 68.18 & 100.00 & 66.67 & 99.64 & 68.49 & 83.05 & 71.00 \\
MANU & 100.00 & 68.18 & 100.00 & 66.67 & 99.64 & 68.72 & 83.05 & 70.62 \\
MMUN-Both & 16.67 & 9.09 & 87.50 & 22.22 & 95.00 & 65.56 & 85.17 & 66.48 \\
MMUN-Lang & 28.57 & 4.55 & 100.00 & 0.00 & 95.84 & 67.18 & 83.26 & 70.62 \\
MMUN-Vis & 23.81 & 9.09 & 87.50 & 0.00 & 94.92 & 68.26 & 83.05 & 68.55 \\
\midrule
\multicolumn{9}{@{}l}{\textit{Forget-5}} \\
\rowcolor[HTML]{F2F5F7}
Base & 13.33 & 29.09 & 100.00 & 62.22 & 9.96 & 22.98 & 100.00 & 51.92 \\
\rowcolor[HTML]{F2F5F7}
Vanilla & 98.57 & 70.00 & 100.00 & 75.56 & 99.74 & 68.18 & 81.82 & 70.71 \\
\midrule
GA & 97.62 & 73.64 & 100.00 & 75.56 & 99.70 & 69.17 & 82.50 & 71.52 \\
GA-Diff & 98.57 & 47.27 & 100.00 & 66.67 & 99.74 & 67.36 & 81.82 & 69.70 \\
KL-Min & 98.57 & 70.00 & 100.00 & 75.56 & 99.74 & 68.26 & 81.82 & 70.30 \\
NPO & 98.57 & 57.27 & 100.00 & 77.78 & 99.74 & 67.77 & 81.82 & 70.10 \\
MANU & 98.57 & 67.27 & 97.50 & 75.56 & 99.39 & 65.04 & 69.32 & 69.70 \\
MMUN-Both & 33.33 & 19.09 & 100.00 & 33.33 & 94.59 & 66.45 & 85.45 & 70.10 \\
MMUN-Lang & 42.38 & 11.82 & 100.00 & 13.33 & 94.46 & 65.54 & 81.82 & 70.10 \\
MMUN-Vis & 25.71 & 9.09 & 100.00 & 13.33 & 93.90 & 65.79 & 81.82 & 67.07 \\
\bottomrule
\end{tabular}
\end{table*}

\begin{table*}[t]
\centering
\footnotesize
\setlength{\tabcolsep}{3.7pt}
\renewcommand{\arraystretch}{1.08}
\caption{\textbf{LLaVA-1.5-13B results under Forget-1 and Forget-5}.}
\label{tab:llava_13b}
\vspace{6pt}
\begin{tabular}{@{}lcccc@{\hspace{5pt}\vrule width 0.35pt\hspace{5pt}}cccc@{}}
\toprule
& \multicolumn{4}{c}{\textbf{Forget residual} $\downarrow$}
& \multicolumn{4}{c}{\textbf{Retain utility} $\uparrow$} \\
\cmidrule(lr){2-5}\cmidrule(l){6-9}
\textbf{Method} & \textbf{AA} & \textbf{IA} & \textbf{IB$^{+}$} & \textbf{IR}
& \textbf{AA} & \textbf{IA} & \textbf{IB$^{+}$} & \textbf{IR} \\
\midrule
\multicolumn{9}{@{}l}{\textit{Forget-1}} \\
\rowcolor[HTML]{F2F5F7}
Base & 11.90 & 31.82 & 75.00 & 66.67 & 9.04 & 25.19 & 90.47 & 54.61 \\
\rowcolor[HTML]{F2F5F7}
Vanilla & 100.00 & 68.18 & 100.00 & 77.78 & 99.60 & 68.34 & 83.05 & 58.76 \\
\midrule
GA-Diff & 100.00 & 50.00 & 100.00 & 77.78 & 97.34 & 67.33 & 83.69 & 64.60 \\
KL-Min & 100.00 & 68.18 & 100.00 & 77.78 & 99.60 & 68.57 & 83.05 & 60.26 \\
NPO & 92.86 & 68.18 & 100.00 & 77.78 & 99.64 & 68.72 & 83.05 & 60.08 \\
MANU & 100.00 & 68.18 & 100.00 & 77.78 & 99.60 & 68.49 & 83.05 & 58.57 \\
MMUN-Both & 23.81 & 9.09 & 100.00 & 22.22 & 94.03 & 65.25 & 94.28 & 58.19 \\
MMUN-Lang & 69.05 & 18.18 & 100.00 & 11.11 & 98.47 & 65.95 & 86.86 & 65.73 \\
MMUN-Vis & 35.71 & 9.09 & 100.00 & 22.22 & 93.91 & 61.71 & 100.00 & 61.02 \\
\midrule
\multicolumn{9}{@{}l}{\textit{Forget-5}} \\
\rowcolor[HTML]{F2F5F7}
Base & 10.48 & 30.00 & 95.00 & 66.67 & 8.96 & 24.88 & 89.77 & 53.74 \\
\rowcolor[HTML]{F2F5F7}
Vanilla & 98.57 & 68.18 & 100.00 & 71.11 & 99.70 & 68.35 & 81.82 & 57.98 \\
\midrule
GA & 54.76 & 68.18 & 100.00 & 53.33 & 66.45 & 66.78 & 100.00 & 50.71 \\
GA-Diff & 82.86 & 61.82 & 100.00 & 68.89 & 98.74 & 66.69 & 81.82 & 55.56 \\
KL-Min & 98.57 & 68.18 & 100.00 & 71.11 & 99.70 & 68.68 & 81.82 & 59.19 \\
NPO & 67.62 & 68.18 & 100.00 & 64.44 & 83.94 & 69.01 & 93.86 & 52.93 \\
MANU & 98.57 & 68.18 & 100.00 & 71.11 & 99.74 & 68.26 & 81.82 & 58.59 \\
MMUN-Both & 38.57 & 12.73 & 100.00 & 31.11 & 95.63 & 63.64 & 100.00 & 57.37 \\
MMUN-Lang & 42.38 & 21.82 & 100.00 & 31.11 & 94.55 & 62.48 & 81.82 & 54.34 \\
MMUN-Vis & 60.00 & 10.00 & 100.00 & 20.00 & 96.84 & 66.28 & 82.95 & 65.45 \\
\bottomrule
\end{tabular}
\end{table*}

\subsection{Llama-3.2-11B Results under Forget-1 and Forget-5}
\label{app:llama_results}

Table~\ref{tab:llama_11b} reports the
complete Forget-1 and Forget-5 results for Llama-3.2-11B. We report the four
task families separately on both the forget and retain splits. 

\begin{table*}[t]
\centering
\footnotesize
\setlength{\tabcolsep}{3.7pt}
\renewcommand{\arraystretch}{1.08}
\caption{\textbf{Llama-3.2-11B results under Forget-1 and Forget-5}.}
\label{tab:llama_11b}
\vspace{6pt}
\begin{tabular}{@{}lcccc@{\hspace{5pt}\vrule width 0.35pt\hspace{5pt}}cccc@{}}
\toprule
& \multicolumn{4}{c}{\textbf{Forget residual} $\downarrow$}
& \multicolumn{4}{c}{\textbf{Retain utility} $\uparrow$} \\
\cmidrule(lr){2-5}\cmidrule(l){6-9}
\textbf{Method} & \textbf{AA} & \textbf{IA} & \textbf{IB$^{+}$} & \textbf{IR}
& \textbf{AA} & \textbf{IA} & \textbf{IB$^{+}$} & \textbf{IR} \\
\midrule
\multicolumn{9}{@{}l}{\textit{Forget-1}} \\
\rowcolor[HTML]{F2F5F7}
Base & 0.00 & 9.09 & 0.00 & 33.33 & 4.12 & 12.17 & 0.00 & 22.03 \\
\rowcolor[HTML]{F2F5F7}
Vanilla & 100.00 & 59.09 & 0.00 & 88.89 & 99.88 & 62.48 & 0.64 & 81.54 \\
\midrule
GA & 88.10 & 68.18 & 0.00 & 77.78 & 99.88 & 63.48 & 0.21 & 81.17 \\
GA-Diff & 100.00 & 59.09 & 0.00 & 88.89 & 99.39 & 60.86 & 0.00 & 78.34 \\
KL-Min & 100.00 & 59.09 & 0.00 & 88.89 & 99.88 & 61.86 & 0.00 & 80.41 \\
NPO & 100.00 & 59.09 & 0.00 & 77.78 & 99.88 & 62.10 & 0.42 & 81.36 \\
MANU & 100.00 & 59.09 & 0.00 & 88.89 & 99.72 & 60.55 & 0.64 & 79.10 \\
R2MU & 7.14 & 22.73 & 12.50 & 44.44 & 4.88 & 20.11 & 6.78 & 38.61 \\
\midrule
\multicolumn{9}{@{}l}{\textit{Forget-5}} \\
\rowcolor[HTML]{F2F5F7}
Base & 2.86 & 8.18 & 0.00 & 20.00 & 4.16 & 12.48 & 0.00 & 22.42 \\
\rowcolor[HTML]{F2F5F7}
Vanilla & 98.57 & 56.36 & 2.50 & 82.22 & 100.00 & 62.98 & 0.45 & 81.62 \\
\midrule
GA & 58.10 & 74.55 & 0.00 & 77.78 & 90.48 & 63.80 & 0.00 & 77.98 \\
GA-Diff & 89.52 & 47.27 & 0.00 & 77.78 & 99.18 & 55.04 & 0.91 & 71.52 \\
KL-Min & 98.57 & 56.36 & 0.00 & 82.22 & 100.00 & 60.66 & 0.00 & 81.41 \\
NPO & 98.10 & 60.91 & 0.00 & 82.22 & 99.91 & 63.80 & 1.14 & 82.02 \\
MANU & 96.19 & 55.45 & 0.00 & 82.22 & 99.57 & 60.66 & 1.59 & 79.19 \\
R2MU & 8.10 & 20.91 & 5.00 & 28.89 & 4.63 & 20.08 & 7.05 & 39.60 \\
\bottomrule
\end{tabular}
\end{table*}

\section{Extended Analysis and Discussion}
\label{app:extended_analysis}
\noindent\textbf{Masked evidence remains a substantial identity cue.}
Among Vanilla-correct queries, the average residual is $53.88\%$ for direct
identity access and $51.85\%$ for indirect identity access across the valid
MMUN configurations. On Qwen3-VL-4B, MMUN-Lang retains $70.00\%$ direct IA
and $70.37\%$ indirect IA. Removing a name or explicit identifier therefore
changes the cue available to the model without necessarily removing the
identity signal carried by the document layout and surrounding attributes.

\noindent\textbf{Biometric continuity is more persistent than record linkage.}
Vanilla-conditioned recovery averages $94.00\%$ for cross-instance binding and
$81.80\%$ for cross-modal binding over the valid MMUN configurations. All
three MMUN variants preserve every eligible cross-instance binding on
LLaVA-1.5-7B, LLaVA-1.5-13B, Qwen3-VL-2B, and Qwen3-VL-4B. On Qwen3-VL-8B,
MMUN-Vis removes all eligible cross-modal bindings while $40.00\%$ of
cross-instance bindings remain recoverable. Since IB$^{+}$ contains matched
positive pairs, these values measure persistence of stored bindings rather than
balanced same/different classification.

\noindent\textbf{Relational composition can recover an identity after direct
access weakens.}
GA-Diff on Qwen3-VL-2B reduces forget AA to $44.76\%$, while forget IR
remains $73.33\%$. MMUN-Lang on Qwen3-VL-4B shows the same separation, with
$44.76\%$ AA and $75.56\%$ IR. More specifically, it retains every
Vanilla-correct relational and relational-attribute reconstruction query on
that model. Residual relations can therefore provide an access route even when
an isolated attribute query has been suppressed.

\noindent\textbf{Additional attributes do not guarantee stronger
reconstruction.}
Across the MMUN configurations, Vanilla-conditioned residuals are $44.30\%$
for relational reconstruction and $45.60\%$ for relational-attribute
reconstruction. The additional attribute can disambiguate a surviving relation,
but it can also be unavailable after unlearning. Cross-modal relational
reconstruction has a lower Vanilla acquisition rate, ranging from $13.33\%$
to $33.33\%$ across backbones. Its raw residual should therefore be
interpreted together with the number of Vanilla-correct examples.

\noindent\textbf{Person--attribute edges are removed asymmetrically.}
For MMUN-Lang, attribute-to-identity recovery on Qwen3-VL-2B, 4B, and 8B
is $66.67\%$, $100.00\%$, and $60.00\%$, whereas identity-to-attribute
recovery is $16.67\%$, $40.00\%$, and $4.00\%$. On LLaVA-1.5-7B, reverse
access is removed for every eligible pair while $20.00\%$ retain only forward
access. The surviving lookup direction depends on the backbone, so one query
direction cannot certify deletion of the underlying association.

\noindent\textbf{Textual and visual access can fail in opposite ways.}
On Qwen3-VL-2B, MMUN-Lang retains $24.64\%$ of Vanilla-correct textual
attribute queries and $47.83\%$ of visual queries. On LLaVA-1.5-13B,
MMUN-Vis retains $73.91\%$ textual queries and $54.35\%$ visual queries.
The weaker modality path is therefore not fixed by the nominal modality of the
unlearning objective. Shared multimodal representations can move residual
access to another input route.

\noindent\textbf{The effective modality target depends on the backbone.}
MMUN-Vis gives the lowest any-path recovery on LLaVA-1.5-7B and
Qwen3-VL-2B, at $42.03\%$ and $57.97\%$. It becomes the weakest MMUN
variant on LLaVA-1.5-13B, where recovery reaches $82.61\%$. MMUN-Lang is
strongest on LLaVA-1.5-13B, Qwen3-VL-4B, and Qwen3-VL-8B, with recovery rates
of $65.22\%$, $66.67\%$, and $24.64\%$. A fixed language or vision mask does
not identify a universal location of identity information.

\noindent\textbf{Deletion load interacts with the backbone.}
For MMUN-Lang, increasing the forget set from one to five identities raises
any-path recovery from $42.86\%$ to $62.32\%$ on LLaVA-1.5-7B, from
$21.43\%$ to $71.01\%$ on Qwen3-VL-2B, and from $35.71\%$ to $66.67\%$ on
Qwen3-VL-4B. The trend reverses on LLaVA-1.5-13B and Qwen3-VL-8B, where
recovery falls from $92.86\%$ to $65.22\%$ and from $35.71\%$ to $24.64\%$.
Forget-1 is therefore a single-identity diagnostic, not a population-level
estimate of multi-identity deletion.

\noindent\textbf{Method rankings do not transfer reliably across backbones.}
GA leaves forget AA at $97.62\%$ on LLaVA-1.5-7B, but reduces it to
$54.76\%$ on LLaVA-1.5-13B and $58.10\%$ on Llama-3.2-11B. Their retain AA
values are $66.45\%$ and $90.48\%$ for the latter two models. GA-Diff reaches
$44.76\%$ forget AA on Qwen3-VL-2B but $98.57\%$ on LLaVA-1.5-7B. An
unlearning objective can therefore appear ineffective, selective, or
destructive depending on the backbone.

\noindent\textbf{Alternative identity keys create disagreement rather than a
single dominant path.}
Under MMUN-Lang on Qwen3-VL-2B, name, face, and fingerprint keys recover
$24.64\%$, $55.07\%$, and $40.58\%$ of eligible attributes individually,
while their union recovers $71.01\%$. At 4B, the strongest individual path
recovers $53.62\%$, but the union reaches $66.67\%$. At 8B, every individual
path is below $16\%$, yet the union remains $24.64\%$. Different keys disagree
on $20.29$--$56.52\%$ of eligible cases. A low residual for one identity key
does not establish that the attribute is inaccessible through another key.

\noindent\textbf{The benchmark exposes complementary failure modes rather than
one scalar notion of forgetting.}
Direct attribute access, identity access, binding, and reconstruction can move
in different directions under the same update. A method can suppress a named
answer while preserving a reverse lookup, a biometric link, or a relational
reconstruction route. The full task profile is therefore necessary for
interpreting whether an individual has been removed as a connected identity.

\section{Limitations and Future Directions}
\label{app:limitations_future}
\textbf{Synthetic profiles and external validity.}
IDUnlearn-Bench uses synthetic individual profiles to provide controlled identity--attribute associations, cross-modal evidence, and reproducible forgetting targets without exposing real personal information. While this design enables precise construction and auditing of individual-level evidence, synthetic profiles cannot fully capture the diversity, noise, and ambiguity of naturally occurring personal data. The current benchmark therefore evaluates individual-level unlearning under a controlled setting rather than claiming to represent all real-world identity distributions. Future versions could extend the benchmark to broader linguistic, cultural, demographic, and temporal settings, as well as less structured evidence distributions.

\textbf{Behavioral scope of the evaluation.}
Our evaluation measures whether target information remains behaviorally recoverable through the access, binding, and reconstruction paths instantiated by the benchmark. A low leakage score indicates that the evaluated queries fail to recover the corresponding information, but does not certify that all internal representations of the target have been removed from the model. As with other behavioral evaluations of machine unlearning, stronger adaptive probing or previously unseen access paths may reveal residual information not captured by the current task suite.

\textbf{Toward stronger auditing and structure-aware unlearning.}
Future work should extend individual-level evaluation beyond fixed single-query tests toward adaptive multi-query recovery, multi-turn interaction, harder cross-modal association tests, and more diverse evidence graphs. More importantly, our results motivate unlearning methods that explicitly account for the structured nature of individual knowledge. Rather than removing isolated attributes alone, future methods could target identity--attribute associations, cross-modal links, and relational evidence while preserving information associated with unrelated individuals. We view such structure-aware unlearning as an important direction exposed by IDUnlearn-Bench, rather than a capability assumed or implemented by the current benchmark.
\end{document}